\documentclass[sigconf]{acmart}
\usepackage{booktabs}
\usepackage{multirow}
\usepackage{graphicx}
\usepackage[normalem]{ulem}
\usepackage{colortbl}
\usepackage{pifont}
\usepackage{kotex}
\usepackage{algorithm}
\usepackage{algpseudocode}
\usepackage{amsmath}
\usepackage{enumitem}
\usepackage{needspace}
\usepackage{makecell}
\usepackage{stackengine}
\usepackage{tabularx}
\usepackage{tcolorbox}
\usepackage{ragged2e}

\usepackage{tikz, xcolor}
\usetikzlibrary{positioning, arrows.meta, calc}

\definecolor{caseblue}{HTML}{0070C0}
\definecolor{casepurple}{HTML}{7030A0}

\DeclareRobustCommand{\redtext}[1]{\textcolor{red}{#1}}
\DeclareRobustCommand{\bluetext}[1]{\textcolor{caseblue}{#1}}
\DeclareRobustCommand{\purpletext}[1]{\textcolor{casepurple}{#1}}

\AtBeginDocument{%
  }

\setcopyright{acmlicensed}
\acmDOI{XXXXXXX.XXXXXXX}
\copyrightyear{2027}
\acmISBN{978-1-4503-XXXX-X/2018/06}

\begin{document}

\title{Federated Recommender Agents}

\author{Jaehyung Lim}
\affiliation{%
  \institution{Pohang University of\\ Science and Technology}
  \city{Pohang}
  \country{Republic of Korea}}
\email{jaehyunglim@postech.ac.kr}

\author{Woojoo Kim}
\affiliation{%
  \institution{Pohang University of\\ Science and Technology}
  \city{Pohang}
  \country{Republic of Korea}}
\email{kimuj0103@postech.ac.kr}

\author{Junyoung Kim}
\affiliation{%
  \institution{Pohang University of\\ Science and Technology}
  \city{Pohang}
  \country{Republic of Korea}}
\email{junyoungkim@postech.ac.kr}

\author{Dongha Kim}
\affiliation{%
  \institution{Pohang University of\\ Science and Technology}
  \city{Pohang}
  \country{Republic of Korea}}
\email{dhkim0317@postech.ac.kr}

\author{Wonbin Kweon}
\authornote{Corresponding authors.}
\affiliation{%
  \institution{Sungkyunkwan University}
  \city{Suwon}
  \country{Republic of Korea}}
\email{wonbinkweon@skku.edu}

\author{Hwanjo Yu}
\authornotemark[1]
\affiliation{%
  \institution{Pohang University of Science and Technology}
  \city{Pohang}
  \country{Republic of Korea}}
\email{hwanjoyu@postech.ac.kr}




\begin{abstract}
LLM-powered recommender agents use reasoning and memory
for personalization, but centralized access to user
information raises privacy concerns, while isolated agents
cannot learn from others' experiences.
We formulate \emph{Federated Recommender Agents},
a setting where local agents collaborate through
natural-language knowledge while keeping interaction
histories and personal memories local.
We propose \textbf{FedRecAgent}, which enables this
collaboration without updating LLM parameters.
Its central challenge is determining which reasoning
to integrate, since similar behavioral contexts can
lead to different choices.
FedRecAgent represents local experiences as
Condition--Transition--Reasoning (C--T--R) units and
hierarchically integrates reasoning by condition
and transition.
Within this structure, it consolidates related
explanations and discriminates reasoning across
alternative transitions.
Each client maintains semantic and decision memories
to capture preferences and conditional choice tendencies,
guiding the retrieval and selection of shared reasoning
for recommendation.
Experiments across three domains under dense and sparse
settings demonstrate improved recommendation performance
over federated and LLM-based baselines, while ablations
support the contributions of personalized memories
and structured reasoning integration.
\end{abstract}

\ccsdesc[500]{Information systems~Retrieval models and ranking}
\ccsdesc[500]{Information systems~Recommender systems}
\ccsdesc[500]{Computing methodologies~Distributed artificial intelligence}

\keywords{Recommender Agents, Federated Recommendation, Federated Recommender Agents}


\maketitle

\section{Introduction}
\label{sec:introduction}

Recently, Large Language Model (LLM)-powered recommender agents have emerged as a promising approach to personalized recommendation by leveraging pretrained knowledge and reasoning capabilities to interpret user interactions \cite{agentcf,agentcf++,agentdr,i2agent,macf,gimo,starec,memrec,recmind,afl}.
These recommender agents can capture user preferences and behavioral contexts from past experiences and utilize the accumulated knowledge (e.g., memory) to inform subsequent recommendation decisions.
However, such personalization often requires personal information, including user profiles, interaction histories, and preferences, raising privacy concerns when this information is processed by centralized servers or external LLM services.
These concerns are further underscored by privacy regulations, such as GDPR~\cite{gdpr} and CCPA~\cite{ccpa}, which impose restrictions on the collection, processing, and sharing of personal information.

A naive remedy is to deploy recommender agents independently on local devices to keep personal information private.
However, this approach limits their ability to leverage diverse behavioral patterns and decision-making experiences observed from other users.
Therefore, a collaborative recommendation framework that keeps personal information local while enabling recommender agents to leverage knowledge derived from one another's experiences is needed.
From this perspective, Federated Recommendation (FedRec) is proposed as a promising approach to enabling collaboration across local devices while keeping user data private \cite{fedmf, fedncf, pfedrec, fedrap, fedem}.
Existing FedRec methods have enabled such collaboration
through numerical aggregation of model
parameters~\cite{pfedrec,gpfedrec,fedrap,cofedrec,fedem}
or collaborative signals~\cite{fedrecgrad,fedcia}.

\begin{figure}
    \centering
    \includegraphics[width=1\linewidth]{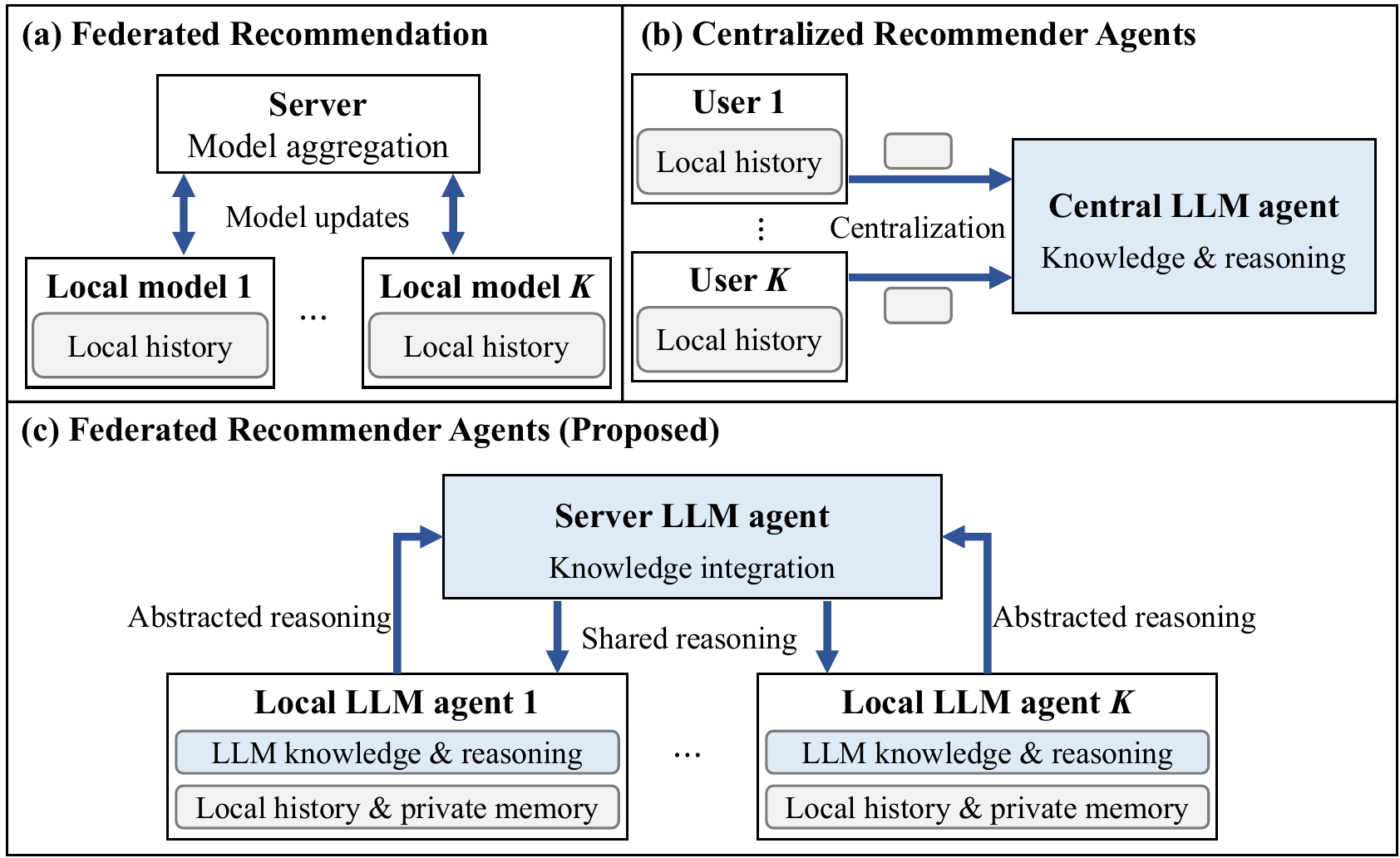}
    \caption{A conceptual comparison between (a) Federated Recommendation, (b) Centralized Recommender Agents, and (c) Federated Recommender Agents (proposed).}
    \label{fig:intro}
\end{figure}

However, FedRec clients typically operate on resource-constrained personal devices, and repeatedly updating the parameters of LLM-based agents can incur substantial computational and memory overhead.
Meanwhile, recent recommender agents can accumulate and update \textit{natural-language knowledge} derived from user experiences without modifying model parameters.
This capability suggests that recommender agents can adapt not only through parameter updates but also by explicitly maintaining and utilizing experience-derived knowledge.
By extending this approach, we envision a new paradigm for FedRec in which agents share natural-language knowledge rather than model parameters, enabling collaborative learning without repeated LLM fine-tuning (Figure~\ref{fig:intro}).
We refer to this new research setting as \textbf{Federated Recommender Agents}.

Realizing such language-based knowledge collaboration between the federated recommender agents introduces three key challenges: (\textbf{C1}) continuously accumulating personalized knowledge, (\textbf{C2}) extracting shareable knowledge, and (\textbf{C3}) integrating knowledge generated by different agents.\\
\textbf{C1}: First, local agents must continuously accumulate and utilize knowledge from new user experiences without updating LLM parameters. In particular, they need to maintain personalized knowledge that captures not only what users prefer but also how they tend to make decisions across different behavioral contexts. 

\noindent \textbf{C2}: Second, agents must determine how to represent
knowledge extracted from local experiences so that other
agents can interpret and reuse it without directly accessing
user interaction histories or specific preferences.
Unlike model parameters, recommendation knowledge expressed
in natural language lacks predefined shared identifiers
for alignment.
We therefore need to define a consistently structured
\emph{unit} of knowledge sharing that makes explicit
what each contribution describes.
For recommendation, such a unit should capture patterns
in the user's history and explain why those
interactions may lead to the observed next choice, while
abstracting away specific item identities and user-specific
preference details.

\noindent \textbf{C3}: Third, we need to establish \emph{criteria for how
these knowledge units should be integrated}.
Recommendation knowledge describes how users with particular
interaction patterns proceed to their next choices.
However, similar patterns may lead to different decisions,
such as continuing familiar interests or exploring new
directions. A common representation alone therefore does
not determine which contributions should be combined.
Integration must account for both the preceding interaction
patterns and the subsequent decision directions, consolidating
related reasoning while preserving explanations for
distinct choices.

To address these challenges, we propose \textbf{FedRecAgent}, a structured reasoning federation framework for Federated Recommender Agents.
FedRecAgent keeps both client-side and server-side LLM parameters fixed, enabling personalization and collaboration through updates to experience-derived natural-language knowledge.
Specifically, each client manages shareable knowledge units and private memories from local experiences.
Each shareable unit follows a Condition--Transition--Reasoning (C--T--R) structure: Condition describes the behavioral context, Transition captures the observed decision transition, and Reasoning explains the corresponding choice. These units abstract away specific item identities and user-specific preference details.
Meanwhile, each local agent maintains two complementary
private memories on its device: \textit{semantic memory}, capturing user preferences, and \textit{decision memory}, capturing choice tendencies across behavioral contexts. The server organizes uploaded units by condition and transition and integrates their reasoning into a hierarchical global knowledge bank. During recommendation, each client uses its private memories to retrieve and select shared reasoning from the bank for personalized ranking.

The main contributions of this paper are summarized as follows:
\begin{itemize}[leftmargin=*]
    \item \textbf{Federated Recommender Agents.}
    We formulate a new research problem, termed \emph{Federated Recommender Agents}, in which LLM-powered recommender agents collaboratively leverage knowledge acquired from other agents while keeping private user information local.
    To address this problem, we propose \textbf{FedRecAgent}, a framework that enables federated collaboration through the sharing and integration of natural-language knowledge rather than model parameters.

    \vspace{+0.1cm}
    \item \textbf{Structured Reasoning Federation.}
    In FedRecAgent, we design private dual memories to accumulate personalized knowledge and formulate C--T--R as a consistent unit for knowledge sharing that abstracts away private user and item details.
    We further establish condition--transition alignment as the criterion for hierarchical reasoning integration.

    \vspace{+0.1cm}
    \item \textbf{Experiments.}
    We evaluate FedRecAgent across three recommendation domains
    under both dense and sparse settings, comparing it with
    representative federated and LLM-based recommendation methods.
    Ablation studies and additional analyses further examine
    the effectiveness of the proposed components.
\end{itemize}
\section{Related Work}

\noindent \textbf{Federated Recommendation.}
Federated recommendation (FedRec) enables collaborative preference learning
while keeping user interaction data on local devices.
Early approaches primarily relied on federated optimization of collaborative
filtering models by sharing model parameters or item representations
instead of raw interaction histories~\cite{fedmf,pfedrec}.
Subsequent studies have explored personalized model adaptation
~\cite{gpfedrec, fedrap, fediar, fedem}, collaborative information aggregation
~\cite{cofedrec,fedcia}, continual preference learning~\cite{fcrec, fcrec2} and cold-start recommendation~\cite{fedcolduser,ifedrec}.
Despite these advances, collaboration has largely relied on learned representations or model updates rather than explicit, reusable reasoning knowledge.

Recently, foundation models have been incorporated into FedRec to leverage their semantic understanding of users and items. \cite{gptfedrec} combines federated retrieval with LLM-based reranking, while \cite{fellrec} introduces an efficient framework for federated LLM-based recommendation. \cite{fedecider} explores personalized LLM adaptation for federated cross-domain recommendation, whereas \cite{fedvlr} and \cite{mmfedrec} leverage pretrained vision-language representations and personalized multimodal fusion. Although these approaches incorporate knowledge, they primarily focus on recommendation modeling, representation fusion, or model adaptation rather than federating reasoning knowledge derived from personalized agent experiences.

\vspace{+0.1cm}
\noindent \textbf{LLM-based Recommender Agents.}
Recent advances in LLMs have enabled recommender agents that
leverage reasoning, planning, and tool utilization for personalized
recommendation. \cite{recmind} employs autonomous planning and tool use, while \cite{agentcf} models users and items as interacting agents to capture collaborative preferences through reflection. Other studies investigate deliberative reasoning~\cite{starec}, dynamic item relationships~\cite{agentdr}, and multi-agent collaboration~\cite{macf}.

Recent studies have also emphasized persistent memory and personalized adaptation. \cite{agentcf++} incorporates memory for cross-domain recommendation, \cite{afl} models iterative feedback between users and recommendation agents, and \cite{i2agent} maintains user-specific memory to adapt to individual instructions and feedback. \cite{memrec} further introduces collaborative memory to capture relational information across users and items. \cite{drunkagent} examines how adversarial textual perturbations can corrupt agent memories and manipulate recommendations. 
However, these studies do not primarily address how recommendation agents can share reusable reasoning while keeping user histories and personal memories local. This requires separating knowledge intended for cross-user collaboration from private personalization information.

\vspace{+0.1cm}
\noindent \textbf{LLM-based Federated Agents.}
Recent studies have extended federated learning to LLM-based agents, enabling collaboration without centralizing local experiences. Some approaches optimize agent models or policies through parameter-based federation~\cite{fedagent,fedwave}.
In contrast, knowledge-based approaches exchange natural-language knowledge instead of model parameters.
\cite{fical} aggregates tool-use knowledge extracted from local data, while \cite{fedtextgrad} uses textual feedback to optimize local prompts and aggregates them through server-side summarization.
Other studies explore federated reasoning insights~\cite{fot}, reusable skills~\cite{federatedskill}, and semantic knowledge abstractions~\cite{fedagentke}. \cite{fedworld} further studies the applicability of shared transition rules across heterogeneous agent environments. These studies demonstrate the potential of knowledge-level collaboration among LLM-based agents. 

However, federating reasoning knowledge for recommendation requires distinguishing different decision transitions across sequential interactions. Even under similar behavioral contexts, users may continue their current interests, return to previous ones, or explore new alternatives. Thus, contextual similarity alone is insufficient to determine which reasoning contributions should be integrated. Furthermore, utilizing shared reasoning requires accounting for individual preferences and decision-making tendencies. FedRecAgent addresses these requirements by organizing shared reasoning into hierarchical \emph{condition--transition} branches and leveraging local private memories for recommendation.
\section{Problem Formulation}
\label{sec:problem}

We formulate a federated recommender agent setting
in which local LLM agents maintain private user data
and personal natural-language knowledge, while
collaborating through shareable knowledge integrated
by a central server-side LLM agent.

\vspace{+0.1cm}
\noindent \textbf{Notations.} \label{sec:recommendation_task}
Let $\mathcal{U}$ and $\mathcal{I}$ denote the sets of users and items, respectively.
For user $u\in\mathcal{U}$, let $i_{u,t}\in\mathcal{I}$
denote the item at position $t$ in the user's
chronologically ordered interaction sequence.
We define the recent history through position $t$ as $S_{u,t}=\bigl(i_{u,k}\bigr)_{k=\max(1,t-L+1)}^{t}$, which contains the most recent $\min(L,t)$ interactions.
An observed episode is represented as $\mathcal{E}_{u,t}=(S_{u,t},i_{u,t+1})$, where $i_{u,t+1}$ is the observed next item.
Training episodes are formed by sliding the history window
forward one interaction at a time, starting when at least
$h_{\min}$ preceding interactions are available.
At inference time, given the current history $S_{u,t}$ and a candidate set $\mathcal{A}_{u,t}\subseteq\mathcal{I}$, the recommendation agent produces a ranking $\pi_{u,t}=\operatorname{ReRank}(S_{u,t},\mathcal{A}_{u,t})$.
The objective is to rank the user's ground-truth next item as highly as possible.

\vspace{+0.1cm}
\noindent \textbf{Local Recommender Agents.} \label{sec:local_agents}
We adopt a user-level federated setting, where each client corresponds to a single user $u\in\mathcal{U}$.
Each client device hosts a local LLM-based recommender agent that interprets the user's interaction experiences to construct shareable and private natural-language knowledge.
The shareable knowledge is abstracted for reuse across users, and the private knowledge captures user-specific preferences and decision tendencies.
The agent maintains and updates its private knowledge as new experiences become available, and uses it together with shared knowledge from the server for recommendation.
Raw interaction histories and private knowledge remain on the client device. 

\vspace{+0.1cm}
\noindent \textbf{Federated Knowledge Exchange.} \label{sec:federated_setting}
Client agents upload their shareable knowledge to a central server, where a server-side LLM agent integrates contributions from clients into reusable shared knowledge and distributes it back to client agents. Both client and server LLM parameters remain frozen; adaptation occurs through updates to private and shared natural-language knowledge.
Training episodes are grouped into synchronous communication
rounds $r=0,\ldots,R-1$ according to their target timestamps,
using common time intervals across users.
A round can therefore contain multiple successive
sliding-window episodes from each client.
Within each round, clients process these episodes
chronologically, updating private knowledge after each
episode while using the shared knowledge finalized
in the previous round.
The server integrates uploaded knowledge once at the
end of each round.

\begin{figure*}
    \centering
    \includegraphics[width=1.0\linewidth]{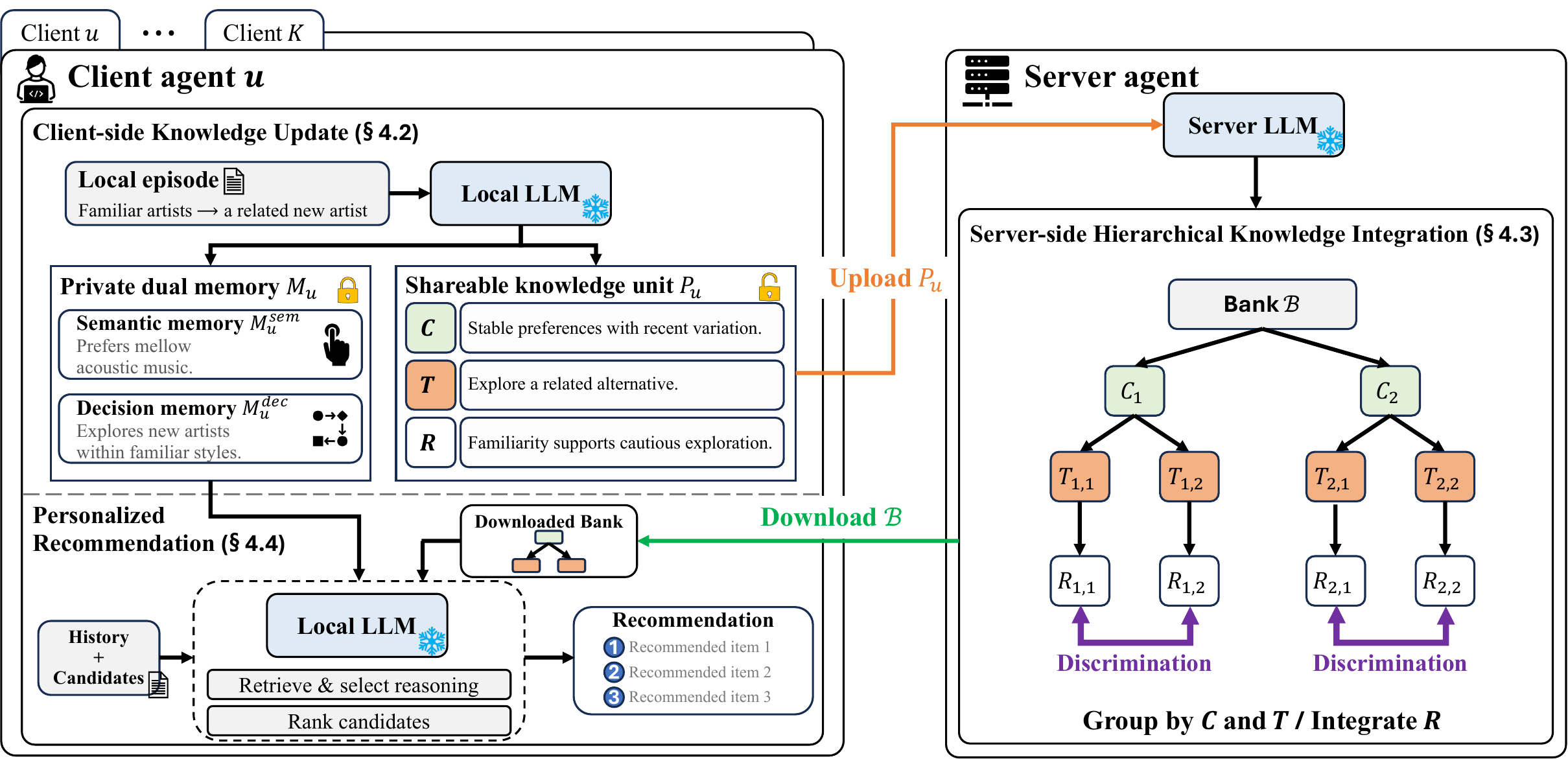}
    \caption{Overview of the FedRecAgent framework.}
    \label{fig:method}
\end{figure*}

\section{Proposed Method: FedRecAgent}
\label{sec:method}
In this section, we first define shareable knowledge units and the global knowledge bank (\S\ref{sec:experience_representation}). We then describe how clients derive these units and update private memories (\S\ref{sec:client_construction}), how the server hierarchically integrates uploaded knowledge (\S\ref{sec:server_integration}), and how clients use shared knowledge and private memories for recommendation (\S\ref{sec:recommendation}).
We use $f_c^{(\cdot)}$ and $f_s^{(\cdot)}$ to denote
client- and server-side LLM calls conditioned on task-specific prompts and instructions, where the superscript identifies the operation.
Figure~\ref{fig:method} illustrates the framework.

\subsection{Knowledge Representation}
\label{sec:experience_representation}
In this section, we define a shareable knowledge unit and describe how these units are organized in the global knowledge bank.

\subsubsection{\textbf{Shareable Knowledge Unit}}
To provide a consistent \emph{unit} of knowledge sharing
without directly transmitting interaction histories, we abstract each local episode $\mathcal{E}_{u,t}$ into $P_{u,t}=(C_{u,t},T_{u,t},R_{u,t})$.
The \textit{condition} $C_{u,t}$ describes the behavioral
context preceding a choice.
The \textit{transition} $T_{u,t}$ describes how the observed
next item relates to preceding interactions.
The \textit{reasoning} $R_{u,t}$ explains why that transition
may occur under the condition.
Generation prompts instruct clients to \emph{omit specific item
identities and user-specific preference details}.

A condition $C$ may describe \textit{interest stability} or
\textit{tension between long-term and recent signals},
while a transition $T$ may express \textit{continuation},
\textit{return}, or \textit{exploration}.
These are illustrative patterns, not predefined categories:
$C$, $T$, and $R$ are free-form natural-language descriptions.
An example is shown below:

\begin{tcolorbox}[
    width=\linewidth,
    colback=gray!8,
    colframe=black,
    boxrule=0.6pt,
    sharp corners,
    boxsep=0pt,
    left=6pt,
    right=6pt,
    top=5pt,
    bottom=5pt,
    before skip=6pt,
    after skip=6pt
]
\small
\justifying
\setlength{\parindent}{0pt}
\setlength{\parskip}{3pt}

\textbf{$C$:} Recent choices consistently reinforce
an established interest.\par
\textbf{$T$:} The next choice continues the same direction.\par
\textbf{$R$:} Repeated engagement may favor continued
exploration within a familiar area rather than a shift
to a different one.\par
\end{tcolorbox}

\subsubsection{\textbf{Hierarchical Knowledge Bank}}
After each communication round $r$, clients
upload their knowledge units to the server.
The server organizes shared knowledge in a hierarchical
bank $\mathcal{B}^{(r)}$: each condition node contains
transition branches, and each branch stores reasoning
for that transition.
We represent this structure as
\begin{equation}
\mathcal{B}^{(r)}
=
\left\{
\left(
\bar{C}_k,\,
\left\{
\left(\bar{T}_{k,j},\bar{R}_{k,j}^{(r)}\right)
\right\}_{j=1}^{J_k^{(r)}}
\right)
\right\}_{k=1}^{K^{(r)}},
\end{equation}
where $K^{(r)}$ is the number of condition nodes and
$J_k^{(r)}$ is the number of transition branches
under condition $k$ after round $r$.
Here, $\bar{C}_k$ describes the condition,
$\bar{T}_{k,j}$ describes its $j$-th transition,
and $\bar{R}_{k,j}^{(r)}$ stores the corresponding
integrated reasoning.
Bars denote stored shared representations, not
arithmetic means.
This organization allows different decision transitions
and their explanations to coexist under the same condition.

\subsection{Client-side Knowledge Update}
\label{sec:client_construction}

In this section, we describe how clients construct a shareable knowledge unit $P_{u,t}$ and update private memories $M_{u,t}$ with episode $\mathcal{E}_{u,t}$. Each client processes local episodes using the shared bank
$\mathcal{B}^{(r-1)}$, finalized in the previous round and distributed by the server. The bank remains fixed throughout the round and is initially empty before the first round.

\subsubsection{\textbf{Shareable Knowledge Derivation}}
\label{sec:local_reasoning_update}

The client first extracts the condition from the preceding
history and the transition from the current episode $\mathcal{E}_{u,t}$:
\begin{equation}
\begin{aligned}
C_{u,t} &= f_c^{\mathrm{condition}}(S_{u,t}),\\
T_{u,t} &= f_c^{\mathrm{transition}}(\mathcal{E}_{u,t}).
\end{aligned}
\label{eq:experience_extraction}
\end{equation}
Condition generation excludes $i_{u,t+1}$ to characterize the behavioral context before the choice is observed, allowing the same operation at inference time. Transition generation uses the observed next item to characterize the direction actually taken.
The client also extracts a compact evidence summary $Z_{u,t}=f_c^{\mathrm{evidence}}(\mathcal{E}_{u,t})$, covering historical preference signals, item characteristics, and their relationships. This episode-specific evidence remains local and grounds reasoning generation and memory updates.

In conventional FedRec, global parameters
provide a starting point for local training \cite{pfedrec,fedrap,fedmf,fedncf}.
We apply an analogous principle at the reasoning level:
clients revise shared reasoning using new local experience,
while keeping LLM parameters frozen.
For each episode, the client retrieves a branch matching
both $C_{u,t}$ and $T_{u,t}$.
The matching procedure is specified in
\S~\ref{sec:hierarchy}.
If no branch matches, including in the initial round
$r=0$ when $\mathcal{B}^{(-1)}=\varnothing$,
the client generates new reasoning:
\begin{equation}
R_{u,t}=f_c^{\mathrm{induce}}
(C_{u,t},T_{u,t},Z_{u,t}).
\label{eq:reasoning_induction}
\end{equation}
Otherwise, the shared reasoning of the matched branch
$(k,j)$ provides the starting point for a local revision:
\begin{equation}
R_{u,t}=f_c^{\mathrm{update}}\!\left(
\bar{R}_{k,j}^{(r-1)},\mathcal{E}_{u,t},Z_{u,t}
\right).
\label{eq:reasoning_update}
\end{equation}
The client preserves supported claims and revises those
unsupported by the current experience.
It uploads the resulting $P_{u,t}$ for server-side
integration, keeping the episode and evidence local.

\subsubsection{\textbf{Private Dual Memory}}
\label{sec:personal_memory}
FedRecAgent distinguishes shared knowledge in the global knowledge bank from private memories maintained locally for each user.
The private memories updated using $\mathcal{E}_{u,t}$ comprise two components: 
\textit{semantic memory} $M_{u,t}^{\mathrm{sem}}$, which captures user preferences, and \textit{decision memory} $M_{u,t}^{\mathrm{dec}}$, which captures choice tendencies under different conditions. 
We denote them by $M_{u,t}=(M_{u,t}^{\mathrm{sem}},M_{u,t}^{\mathrm{dec}})$. 

Each episode $\mathcal{E}_{u,t}$ produces observations for both memories:
\begin{equation}
\begin{aligned}
V_{u,t}^{\mathrm{sem}}
&=f_c^{\mathrm{sem\text{-}obs}}
(\mathcal{E}_{u,t},Z_{u,t}),\\
V_{u,t}^{\mathrm{dec}}
&=f_c^{\mathrm{dec\text{-}obs}}
(\mathcal{E}_{u,t},Z_{u,t},P_{u,t}).
\end{aligned}
\label{eq:memory_observations}
\end{equation}
Semantic observation uses the episode and its evidence.
Decision observation additionally uses $P_{u,t}$ to
interpret the user's choice in relation to the abstract
decision pattern.
The client updates each memory with its observation:
\begin{equation}
M_{u,t}^{q}
=f_c^{q\text{-}\mathrm{mem}}
(M_{u,t-1}^{q},V_{u,t}^{q}),
\quad q\in\{\mathrm{sem},\mathrm{dec}\}.
\label{eq:memory_update}
\end{equation}
The memories are maintained separately to preserve their
complementary roles.
Both remain on the client and complement the global
knowledge bank during recommendation (\S~\ref{sec:recommendation}).

\subsection{Server-side Hierarchical Knowledge Integration}
\label{sec:server_integration}
At the end of communication round $r$, the server integrates the knowledge units uploaded by clients during that round.
A common C--T--R representation does not by itself determine which contributions should be combined. 
We use condition and transition similarity as the \emph{criteria for knowledge integration}.
The server first aligns similar conditions, then groups contributions by transition within each condition, and consolidates reasoning within the resulting branches. 
This hierarchical process preserves distinct decision directions under similar conditions and updates the bank from $\mathcal{B}^{(r-1)}$ to $\mathcal{B}^{(r)}$.

\subsubsection{\textbf{Condition--Transition Assignment}}
\label{sec:hierarchy}
For each uploaded knowledge unit $P_{u,t}=(C_{u,t},T_{u,t},R_{u,t})$, the server
first finds the closest condition prototype using a fixed
sentence encoder $e(\cdot)$:
\begin{equation}
k^*=\operatorname*{arg\,max}_{k}
\cos\!\left(e(C_{u,t}),e(\bar{C}_k)\right).
\label{eq:condition_matching}
\end{equation}
If no condition exists or the highest similarity is below
$\tau_C$, a new condition node is created.
Otherwise, the server searches for a matching transition
within the selected condition:
\begin{equation}
j^*=\operatorname*{arg\,max}_{j}
\cos\!\left(e(T_{u,t}),e(\bar{T}_{k^*,j})\right).
\label{eq:transition_matching}
\end{equation}
A new branch is created if the highest transition similarity
is below $\tau_T$.
A newly created condition starts with a branch initialized
from the uploaded transition.
Condition and transition prototypes remain fixed after
creation.

\subsubsection{\textbf{Within-branch Knowledge Integration}}
\label{sec:within_branch_integration}

After condition--transition assignment, the server
integrates the reasoning associated with each branch
at two temporal scales.
First, \textit{\textbf{distillation}} consolidates reasoning records assigned to the same branch within round $r$.
Let $\mathcal{R}_{k,j}^{(r)}$ denote the collection of reasoning records $R_{u,t}$ uploaded in round $r$ and assigned to transition branch $j$ under condition node $k$.
The server consolidates these records into a single distilled explanation $\widehat{R}_{k,j}^{(r)}$:
\begin{equation}
\widehat{R}_{k,j}^{(r)}
=f_s^{\mathrm{distill}}\!\left(
\bar{C}_k,\bar{T}_{k,j},
\operatorname{Sample}_{B}(\mathcal{R}_{k,j}^{(r)})
\right),
\label{eq:reasoning_distillation}
\end{equation}
where $\operatorname{Sample}_{B}$ selects at most $B$ records.
Distillation compresses redundant explanations while
retaining common decision logic and complementary
applicability conditions.

Second, \textit{\textbf{temporal merging}} combines the distilled
reasoning with the branch's prior rule:
\begin{equation}
\widetilde{R}_{k,j}^{(r)}
=f_s^{\mathrm{merge}}\!\left(
\bar{C}_k,\bar{T}_{k,j},
\bar{R}_{k,j}^{(r-1)},\widehat{R}_{k,j}^{(r)}
\right).
\label{eq:temporal_merging}
\end{equation}
The server is prompted to preserve complementary
explanations and their applicability conditions while
removing unsupported claims.
New branches use $\widehat{R}_{k,j}^{(r)}$ directly;
branches without new uploads retain their previous rules
at this stage.

\subsubsection{\textbf{Cross-branch Reasoning Discrimination}}
\label{sec:branch_refinement}
Although reasoning is integrated within each condition--transition branch, independent integration may yield generic explanations that obscure differences between transitions.
The server therefore compares reasoning across branches under the same condition to clarify the applicability of each branch's reasoning.

Starting from the merged reasoning
$R_{k,j}^{(r,0)}=\widetilde{R}_{k,j}^{(r)}$,
the server iteratively refines each branch's reasoning.
At iteration $h$, let $\mathcal{S}_{k,j}^{(h)}$
denote the indices of at most $Q$ sibling branches
whose current reasoning has the highest cosine
similarity to $R_{k,j}^{(r,h)}$, measured using
the fixed sentence encoder $e(\cdot)$:
\begin{equation}
\begin{aligned}
R_{k,j}^{(r,h+1)}
= f_s^{\mathrm{discriminate}}\!\Big(
\bar{C}_k,\bar{T}_{k,j},R_{k,j}^{(r,h)},
\{(\bar{T}_{k,\ell},R_{k,\ell}^{(r,h)})
\}_{\ell\in\mathcal{S}_{k,j}^{(h)}}
\Big).
\end{aligned}
\label{eq:reasoning_discrimination}
\end{equation}
The sibling transition--reasoning pairs provide
contrasting context for clarifying when the target
reasoning applies to its associated transition.
All branches are updated synchronously using the rules
from iteration $h$, and sibling branches are reselected
using the updated rules for the next iteration.
After $H$ iterations, the final shared reasoning is
$\bar{R}_{k,j}^{(r)}=R_{k,j}^{(r,H)}$.
The updated bank is distributed to clients for the next round.

\newcommand{\faresult}[2]{%
  \mbox{#1\,{\tiny$\pm\!#2$}}%
}

\begin{table*}[t]
  \centering
  \caption{Recommendation performance with random-negative candidates.
    Each cell reports the mean and sample standard deviation over five
    selected seeds.
    Best means are bold and best baseline means are underlined, including ties.}
  \label{tab:main}
  \footnotesize
  \setlength{\tabcolsep}{1.5pt}
  \renewcommand{\arraystretch}{0.95}
  \setlength{\aboverulesep}{0.15ex}
  \setlength{\belowrulesep}{0.15ex}
  \begin{tabularx}{\textwidth}{@{}l*{12}{>{\raggedleft\arraybackslash}X}@{}}
    \toprule
    \multirow{2}{*}{Method}
      & \multicolumn{4}{c}{CDs-Dense} & \multicolumn{4}{c}{Music-Dense} & \multicolumn{4}{c}{Office-Dense} \\
    \cmidrule(lr){2-5}\cmidrule(lr){6-9}\cmidrule(lr){10-13}
      & N@1 & N@5 & N@10 & MRR & N@1 & N@5 & N@10 & MRR & N@1 & N@5 & N@10 & MRR \\
    \midrule
    PFedRec
      & \faresult{9.60}{3.05} & \faresult{29.23}{1.95} & \faresult{45.22}{1.31} & \faresult{29.00}{1.71}
      & \faresult{9.80}{3.49} & \faresult{29.03}{2.89} & \faresult{45.18}{2.30} & \faresult{28.99}{2.96}
      & \faresult{8.80}{2.77} & \faresult{29.29}{3.40} & \faresult{44.83}{1.93} & \faresult{28.44}{2.43} \\
    GPFedRec
      & \faresult{8.00}{2.92} & \faresult{26.76}{5.68} & \faresult{43.63}{2.92} & \faresult{27.02}{3.59}
      & \faresult{8.60}{1.95} & \faresult{28.62}{2.89} & \faresult{44.95}{1.72} & \faresult{28.64}{2.14}
      & \faresult{7.80}{3.11} & \faresult{28.51}{4.64} & \faresult{44.17}{2.67} & \faresult{27.63}{3.35} \\
    FedRAP
      & \faresult{11.00}{3.16} & \faresult{29.29}{2.55} & \faresult{45.62}{1.38} & \faresult{29.55}{1.83}
      & \faresult{10.00}{2.55} & \faresult{30.93}{2.68} & \faresult{45.93}{1.81} & \faresult{29.84}{2.28}
      & \faresult{10.00}{4.64} & \faresult{28.54}{6.55} & \faresult{44.94}{3.44} & \faresult{28.72}{4.30} \\
    FedCIA
      & \faresult{12.20}{1.48} & \faresult{34.58}{1.07} & \faresult{48.02}{0.55} & \faresult{32.47}{0.74}
      & \faresult{10.00}{1.87} & \faresult{29.92}{2.49} & \faresult{45.63}{1.33} & \faresult{29.47}{1.68}
      & \faresult{10.60}{1.82} & \faresult{31.61}{1.40} & \faresult{46.31}{0.68} & \faresult{30.31}{0.91} \\
    \midrule
    FedGRU4Rec
      & \faresult{11.80}{3.27} & \faresult{29.36}{4.43} & \faresult{46.04}{3.05} & \faresult{30.10}{3.81}
      & \faresult{9.40}{3.29} & \faresult{28.05}{1.33} & \faresult{44.50}{1.19} & \faresult{28.13}{1.61}
      & \faresult{9.00}{3.54} & \faresult{29.19}{3.70} & \faresult{45.41}{2.20} & \faresult{29.19}{2.78} \\
    FedSASRec
      & \faresult{10.20}{3.35} & \faresult{27.25}{3.91} & \faresult{44.52}{2.66} & \faresult{28.25}{3.35}
      & \faresult{12.00}{3.67} & \faresult{31.13}{5.28} & \faresult{46.46}{2.82} & \faresult{30.63}{3.55}
      & \faresult{11.80}{3.83} & \faresult{32.96}{5.19} & \faresult{47.38}{3.30} & \faresult{31.70}{4.15} \\
    FedBERT4Rec
      & \faresult{9.60}{3.21} & \faresult{29.91}{5.04} & \faresult{45.09}{2.58} & \faresult{28.81}{3.22}
      & \faresult{12.80}{1.79} & \faresult{32.19}{2.10} & \faresult{47.43}{0.69} & \faresult{31.82}{0.88}
      & \faresult{10.40}{4.04} & \faresult{30.99}{7.33} & \faresult{46.14}{4.05} & \faresult{30.12}{5.01} \\
    \midrule
    FedLLMRank
      & \faresult{14.80}{2.86} & \faresult{35.68}{2.62} & \faresult{49.28}{1.74} & \faresult{34.16}{2.18}
      & \faresult{\underline{14.80}}{4.21} & \faresult{35.22}{4.62} & \faresult{\underline{49.14}}{3.23} & \faresult{\underline{34.00}}{4.08}
      & \faresult{\underline{15.60}}{2.07} & \faresult{36.94}{1.48} & \faresult{\underline{50.57}}{0.77} & \faresult{\underline{35.79}}{1.02} \\
    FedAgentCF
      & \faresult{14.20}{1.30} & \faresult{37.80}{3.26} & \faresult{50.70}{1.67} & \faresult{35.82}{2.01}
      & \faresult{10.80}{1.64} & \faresult{36.30}{2.75} & \faresult{48.78}{1.67} & \faresult{33.28}{2.07}
      & \faresult{10.60}{2.88} & \faresult{35.97}{2.77} & \faresult{48.70}{1.76} & \faresult{33.15}{2.26} \\
    FedAFL
      & \faresult{\underline{15.60}}{1.52} & \faresult{\underline{38.50}}{0.66} & \faresult{\underline{50.78}}{0.78} & \faresult{\underline{35.96}}{1.04}
      & \faresult{12.60}{3.36} & \faresult{\underline{36.45}}{2.66} & \faresult{49.12}{1.99} & \faresult{33.79}{2.54}
      & \faresult{14.80}{2.68} & \faresult{\underline{37.55}}{3.01} & \faresult{50.21}{1.65} & \faresult{35.25}{2.09} \\
    FedI2Agent
      & \faresult{\underline{15.60}}{4.98} & \faresult{34.50}{6.03} & \faresult{49.04}{3.61} & \faresult{33.92}{4.58}
      & \faresult{13.80}{1.48} & \faresult{34.15}{2.19} & \faresult{48.56}{1.14} & \faresult{33.26}{1.38}
      & \faresult{14.60}{2.88} & \faresult{37.39}{2.81} & \faresult{49.93}{2.08} & \faresult{34.87}{2.64} \\
    \midrule
        \rowcolor{gray!12}
    \textbf{FedRecAgent}$_{{D}}$
      & \faresult{\textbf{21.00}}{3.61}
      & \faresult{\textbf{44.57}}{2.35}
      & \faresult{54.92}{1.77}
      & \faresult{41.18}{2.28}
      & \faresult{20.40}{2.41}
      & \faresult{\textbf{40.20}}{3.03}
      & \faresult{53.27}{2.09}
      & \faresult{39.25}{2.59}
      & \faresult{17.80}{3.27}
      & \faresult{\textbf{42.30}}{2.50}
      & \faresult{52.93}{1.62}
      & \faresult{38.60}{2.05} \\

    \rowcolor{gray!12}
    \textbf{FedRecAgent}$_{{P}}$
      & \faresult{\textbf{21.00}}{3.39}
      & \faresult{44.20}{2.89}
      & \faresult{54.76}{2.09}
      & \faresult{41.02}{2.68}
      & \faresult{\textbf{21.20}}{2.77}
      & \faresult{40.19}{3.52}
      & \faresult{\textbf{53.47}}{2.17}
      & \faresult{\textbf{39.54}}{2.74}
      & \faresult{17.40}{2.70}
      & \faresult{40.42}{1.96}
      & \faresult{52.39}{0.68}
      & \faresult{37.97}{0.87} \\

    \rowcolor{gray!12}
    \textbf{FedRecAgent}$_{{P+D}}$
      & \faresult{20.80}{2.77}
      & \faresult{44.13}{4.64}
      & \faresult{\textbf{55.00}}{2.75}
      & \faresult{\textbf{41.34}}{3.41}
      & \faresult{20.80}{2.77}
      & \faresult{40.19}{3.32}
      & \faresult{53.36}{2.12}
      & \faresult{39.39}{2.68}
      & \faresult{\textbf{18.20}}{3.49}
      & \faresult{41.63}{2.35}
      & \faresult{\textbf{53.17}}{1.42}
      & \faresult{\textbf{38.92}}{1.87} \\

    \midrule
    \addlinespace[1pt]
    \multirow{2}{*}{Method}
      & \multicolumn{4}{c}{CDs-Sparse} & \multicolumn{4}{c}{Music-Sparse} & \multicolumn{4}{c}{Office-Sparse} \\
    \cmidrule(lr){2-5}\cmidrule(lr){6-9}\cmidrule(lr){10-13}
      & N@1 & N@5 & N@10 & MRR & N@1 & N@5 & N@10 & MRR & N@1 & N@5 & N@10 & MRR \\
    \midrule
    PFedRec
      & \faresult{9.80}{4.97} & \faresult{31.08}{3.31} & \faresult{45.66}{2.74} & \faresult{29.53}{3.59}
      & \faresult{9.00}{3.94} & \faresult{27.28}{2.45} & \faresult{44.21}{2.51} & \faresult{27.77}{3.24}
      & \faresult{7.80}{3.27} & \faresult{28.09}{2.19} & \faresult{44.62}{1.57} & \faresult{28.21}{2.11} \\
    GPFedRec
      & \faresult{10.60}{3.85} & \faresult{29.85}{4.27} & \faresult{45.94}{2.95} & \faresult{29.98}{3.78}
      & \faresult{9.20}{1.30} & \faresult{30.01}{2.72} & \faresult{45.14}{1.33} & \faresult{28.89}{1.64}
      & \faresult{10.80}{4.02} & \faresult{31.74}{3.77} & \faresult{46.78}{3.09} & \faresult{30.91}{3.90} \\
    FedRAP
      & \faresult{6.60}{2.70} & \faresult{25.64}{1.27} & \faresult{43.19}{0.81} & \faresult{26.41}{1.11}
      & \faresult{9.60}{4.39} & \faresult{31.74}{5.40} & \faresult{46.37}{2.64} & \faresult{30.36}{3.34}
      & \faresult{7.80}{2.17} & \faresult{27.19}{3.83} & \faresult{43.88}{1.92} & \faresult{27.29}{2.37} \\
    FedCIA
      & \faresult{9.40}{5.50} & \faresult{29.91}{4.03} & \faresult{45.44}{3.41} & \faresult{29.24}{4.42}
      & \faresult{10.00}{3.54} & \faresult{29.50}{3.38} & \faresult{45.31}{2.02} & \faresult{29.14}{2.61}
      & \faresult{10.80}{5.72} & \faresult{29.84}{5.53} & \faresult{45.59}{3.57} & \faresult{29.50}{4.56} \\
    \midrule
    FedGRU4Rec
      & \faresult{12.20}{2.39} & \faresult{30.36}{1.13} & \faresult{46.31}{1.15} & \faresult{30.50}{1.56}
      & \faresult{11.80}{1.79} & \faresult{33.41}{3.43} & \faresult{47.38}{1.87} & \faresult{31.68}{2.28}
      & \faresult{10.60}{1.82} & \faresult{28.74}{1.20} & \faresult{45.31}{0.91} & \faresult{29.22}{1.21} \\
    FedSASRec
      & \faresult{11.60}{3.85} & \faresult{31.27}{3.26} & \faresult{46.80}{2.02} & \faresult{31.02}{2.56}
      & \faresult{12.20}{2.17} & \faresult{32.03}{1.35} & \faresult{47.08}{0.48} & \faresult{31.41}{0.66}
      & \faresult{12.60}{3.91} & \faresult{32.60}{3.19} & \faresult{47.19}{2.38} & \faresult{31.49}{3.05} \\
    FedBERT4Rec
      & \faresult{10.20}{2.17} & \faresult{31.09}{2.54} & \faresult{45.99}{1.72} & \faresult{29.97}{2.16}
      & \faresult{10.40}{4.04} & \faresult{31.65}{2.74} & \faresult{46.38}{2.35} & \faresult{30.41}{3.06}
      & \faresult{9.20}{4.27} & \faresult{28.81}{5.74} & \faresult{45.04}{3.40} & \faresult{28.82}{4.23} \\
    \midrule
    FedLLMRank
      & \faresult{\underline{22.20}}{5.02} & \faresult{\underline{42.77}}{4.00} & \faresult{\underline{54.93}}{2.66} & \faresult{\underline{41.44}}{3.42}
      & \faresult{15.60}{2.70} & \faresult{39.32}{3.54} & \faresult{50.89}{1.72} & \faresult{36.08}{2.19}
      & \faresult{15.60}{3.29} & \faresult{38.92}{3.77} & \faresult{51.30}{2.58} & \faresult{36.62}{3.24} \\
    FedAgentCF
      & \faresult{15.60}{5.41} & \faresult{41.64}{3.95} & \faresult{52.33}{3.08} & \faresult{37.78}{4.00}
      & \faresult{11.20}{2.86} & \faresult{39.13}{2.23} & \faresult{50.09}{1.67} & \faresult{34.76}{2.15}
      & \faresult{10.60}{1.14} & \faresult{39.42}{3.36} & \faresult{50.49}{1.54} & \faresult{35.29}{1.82} \\
    FedAFL
      & \faresult{18.20}{2.39} & \faresult{41.09}{3.01} & \faresult{52.62}{1.56} & \faresult{38.28}{2.03}
      & \faresult{15.60}{2.61} & \faresult{\underline{40.07}}{2.60} & \faresult{\underline{51.38}}{1.76} & \faresult{\underline{36.66}}{2.22}
      & \faresult{\underline{18.60}}{3.97} & \faresult{\underline{42.95}}{2.66} & \faresult{\underline{53.89}}{2.11} & \faresult{\underline{39.87}}{2.77} \\
    FedI2Agent
      & \faresult{16.60}{2.07} & \faresult{36.80}{2.77} & \faresult{50.17}{1.48} & \faresult{35.31}{1.84}
      & \faresult{\underline{16.80}}{2.59} & \faresult{37.08}{3.32} & \faresult{50.94}{1.94} & \faresult{36.28}{2.45}
      & \faresult{17.60}{1.95} & \faresult{37.35}{1.47} & \faresult{50.90}{1.19} & \faresult{36.28}{1.50} \\
    \midrule
        \rowcolor{gray!12}
    \textbf{FedRecAgent}$_{{D}}$
      & \faresult{29.20}{2.68}
      & \faresult{49.56}{2.13}
      & \faresult{59.33}{1.65}
      & \faresult{47.02}{2.11}
      & \faresult{\textbf{18.40}}{1.52}
      & \faresult{44.29}{1.52}
      & \faresult{53.75}{0.70}
      & \faresult{39.60}{0.83}
      & \faresult{21.60}{2.07}
      & \faresult{42.38}{2.55}
      & \faresult{54.18}{1.46}
      & \faresult{40.40}{1.84} \\

    \rowcolor{gray!12}
    \textbf{FedRecAgent}$_{{P}}$
      & \faresult{\textbf{30.00}}{2.55}
      & \faresult{\textbf{50.56}}{3.21}
      & \faresult{\textbf{59.98}}{1.62}
      & \faresult{\textbf{47.84}}{2.05}
      & \faresult{17.20}{2.59}
      & \faresult{\textbf{44.54}}{1.64}
      & \faresult{\textbf{53.86}}{1.04}
      & \faresult{\textbf{39.66}}{1.37}
      & \faresult{22.20}{4.32}
      & \faresult{\textbf{43.77}}{3.40}
      & \faresult{\textbf{54.72}}{2.27}
      & \faresult{\textbf{41.07}}{2.94} \\

    \rowcolor{gray!12}
    \textbf{FedRecAgent}$_{{P+D}}$
      & \faresult{29.20}{1.92}
      & \faresult{49.29}{3.70}
      & \faresult{59.50}{1.82}
      & \faresult{47.23}{2.22}
      & \faresult{17.60}{2.19}
      & \faresult{44.27}{1.60}
      & \faresult{53.69}{0.82}
      & \faresult{39.49}{1.04}
      & \faresult{\textbf{22.40}}{3.21}
      & \faresult{43.47}{1.83}
      & \faresult{54.66}{1.54}
      & \faresult{40.98}{2.02} \\

    \bottomrule
  \end{tabularx}
\end{table*}

\subsection{Recommendation by Local Agents}
\label{sec:recommendation}

The shared bank offers multiple decision directions under
similar conditions.
At inference time, the client retrieves relevant reasoning,
selects a suitable branch, and ranks candidate items using
its private memories.
We consider three inference variants that differ in the
memory used as the retrieval query, while sharing the same
training procedure, branch selector, and ranking mechanism.
We omit the round superscript below and use the final bank
and personal memories.

Since the next item is unknown, the client extracts only
the current condition:
$C_u^*=f_c^{\mathrm{condition}}(S_u)$.
It retrieves the most similar condition node $k^*$.
Unlike training-time matching, inference uses the nearest
condition even when its similarity falls below $\tau_C$,
without creating new nodes.

\subsubsection{\textbf{Memory-based Retrieval Variants}}
Semantic memory describes what the user prefers, whereas
decision memory describes how the user tends to choose.
These provide different perspectives for retrieving shared
reasoning.
We define three queries:
\begin{equation}
\begin{aligned}
q_u^{\mathrm{P}} = M_u^{\mathrm{sem}}, \quad
q_u^{\mathrm{D}} = M_u^{\mathrm{dec}}, \quad
q_u^{\mathrm{P+D}}
= \operatorname{concat}
   (M_u^{\mathrm{sem}},M_u^{\mathrm{dec}}).
\end{aligned}
\label{eq:retrieval_queries}
\end{equation}
We denote these variants by
FedRecAgent$_{P}$,
FedRecAgent$_{D}$, and
FedRecAgent$_{P+D}$.
Here, $\operatorname{concat}$ joins the semantic-memory
text followed by the decision-memory text; the two memories
remain separately maintained.

For each variant, let $q_u$ denote its retrieval query.
To limit selection cost, the client retains at most $b$
branches under $k^*$ by cosine similarity between the query and each branch's shared reasoning:
\begin{equation}
\begin{aligned}
s_{u,j}
&=\operatorname{cos}\!\left(
e(q_u),e(\bar{R}_{k^*,j})
\right),\\
\mathcal{J}_u
&=\operatorname{Top}_{b}
\left(\{(j,s_{u,j})\}_{j=1}^{J_{k^*}}\right),
\end{aligned}
\label{eq:branch_candidates}
\end{equation}
where $J_{k^*}$ is the number of branches with nonempty reasoning
under condition $k^*$, and $\mathcal{J}_u$ contains the
indices of the highest-scoring branches.

\subsubsection{\textbf{Branch Selection and Ranking}}
For all three variants, the client selects one branch using
the history, both private memories
$M_u=(M_u^{\mathrm{sem}},M_u^{\mathrm{dec}})$,
and the retrieved transition--reasoning pairs:
\begin{equation}
\begin{aligned}
j_u^*=f_c^{\mathrm{select}}\!\Big(
S_u,C_u^*,M_u,
\{(\bar{T}_{k^*,j},\bar{R}_{k^*,j})\}_{j\in\mathcal{J}_u}
\Big).
\end{aligned}
\label{eq:branch_selection}
\end{equation}
Candidate items are not provided during branch retrieval
or selection.
The selected transition and reasoning then complement
both memories in reranking the candidate set $\mathcal{A}_u$:
\begin{equation}
\begin{aligned}
\pi_u=f_c^{\mathrm{rank}}\!\Big(
S_u,M_u,C_u^*,
\bar{T}_{k^*,j_u^*},\bar{R}_{k^*,j_u^*},\mathcal{A}_u
\Big).
\end{aligned}
\label{eq:candidate_ranking}
\end{equation}
The variants differ only in retrieval-query
construction; both memories inform branch selection
and final ranking in every variant.
\section{Experiments}
\subsection{Experimental Setup}
We briefly summarize the experiment setup due to limited space. Please refer to Appendix~\ref{appendix:exp_details} for the experimental details.

\subsubsection{\textbf{Datasets}}
We use three real-world datasets from the Amazon Review Data~\cite{amazon_data}: CDs \& Vinyl (CDs), Musical Instruments (Music), and Office Products (Office). These domains are widely used in recommendation and LLM-based agent studies~\cite{agentcf,drunkagent,foundation_rs}.
Following prior user-sampling protocols \cite{agentcf,agentcf++,drunkagent}, we construct dense and sparse subsets for each domain, yielding six datasets.

\subsubsection{\textbf{Evaluation Metrics and Protocol}}
We report NDCG@$K$ (N@$K$) \cite{ndcg}, $K\in\{1,5,10\}$,
and Mean Reciprocal Rank (MRR) \cite{mrr}, scaled by 100.
We use per-user chronological leave-one-out
evaluation~\cite{sasrec,agentcf,gpfedrec,fedrap,llmrank}: the last interaction
is held out for testing, the second-to-last for
validation, and the rest for training.
Each ground-truth item is ranked against nine random negatives, with all ten candidates presented in random order.
Personalized memories and the shared bank use only training data and remain fixed during evaluation.
Unless otherwise specified, results are averaged
over users and five random seeds; the main table
also reports standard deviations across seeds.

\subsubsection{\textbf{Baselines}}
We compare FedRecAgent with 11 baselines in three categories:
(1) \textbf{Conventional federated recommendation:}
PFedRec~\cite{pfedrec}, GPFedRec~\cite{gpfedrec},
FedRAP~\cite{fedrap}, and FedCIA~\cite{fedcia};
(2) \textbf{Federated adaptations of sequential models:}
we adapt GRU4Rec~\cite{gru4rec}, SASRec~\cite{sasrec},
and BERT4Rec~\cite{bert4rec} to the federated setting,
denoting them as FedGRU4Rec, FedSASRec, and FedBERT4Rec;
and (3) \textbf{Federated adaptations of LLM-based methods:}
we select LLMRank~\cite{llmrank}, AgentCF~\cite{agentcf},
AFL~\cite{afl}, and I2Agent~\cite{i2agent}
for their amenability to federated adaptation,
and denote our adaptations as FedLLMRank, FedAgentCF,
FedAFL, and FedI2Agent.

\subsubsection{\textbf{Implementation Details}}
We implement all methods in PyTorch with CUDA on up to
five NVIDIA RTX 4090 GPUs.
The client and server backbones are
\texttt{Qwen3-4B-FP8} and \texttt{Qwen3-8B-FP8}~\cite{qwen3},
abbreviated as Qwen3-4B and Qwen3-8B.
The smaller client follows prior work on deploying
8-bit 3B--4B LLMs in resource-constrained
environments~\cite{4b_validation,4b_validation2,4b_validation3},
while the larger server supports shared knowledge integration.
Client and server prompt batch sizes are 32 and 4, respectively.
We use \texttt{all-MiniLM-L6-v2} for semantic similarity,
evaluating $\tau_C,\tau_T\in\{0.3,0.5,0.7\}$
and retrieval budget $b\in\{1,2,3,4,5\}$.
We set $R=4$ communication rounds, maximum history length
$L=10$, and minimum preceding history $h_{\min}=8$.
The server uses at most $B=10$ reasoning records for distillation and $H=2$ discrimination iterations
with up to $Q=10$.

\begin{table}[t]
    \centering
    \caption{Recommendation performance under FedSASRec-based hard-negative
    candidates.}
    \label{tab:main_hard}
    \scriptsize
    \setlength{\tabcolsep}{3pt}
    \setlength{\aboverulesep}{0pt}
    \setlength{\belowrulesep}{0pt}
    \renewcommand{\arraystretch}{0.9}

    \resizebox{\columnwidth}{!}{%
    \begin{tabular}{llcccccc}
        \toprule
        \multirow{2}{*}{Model}
        & \multirow{2}{*}{Metric}
        & \multicolumn{2}{c}{CDs}
        & \multicolumn{2}{c}{Music}
        & \multicolumn{2}{c}{Office} \\
        \cmidrule(lr){3-4}
        \cmidrule(lr){5-6}
        \cmidrule(lr){7-8}
        & & Dense & Sparse & Dense & Sparse & Dense & Sparse \\
        \midrule

        \multirow{2}{*}{FedLLMRank}
        & N@10
        & 49.63 & 50.27
        & 48.38 & \underline{52.28}
        & 50.03 & 48.18 \\
        & MRR
        & 34.68 & 35.60
        & 33.24 & \underline{37.75}
        & 35.22 & 32.81 \\
        \midrule

        \multirow{2}{*}{FedAgentCF}
        & N@10
        & \underline{50.30} & 51.22
        & 47.83 & 48.41
        & 51.86 & 49.19 \\
        & MRR
        & \underline{35.09} & 36.43
        & 32.44 & 32.76
        & 37.20 & 33.63 \\
        \midrule

        \multirow{2}{*}{FedAFL}
        & N@10
        & 43.83 & 43.86
        & 42.37 & 44.28
        & 45.43 & 46.02 \\
        & MRR
        & 26.91 & 26.90
        & 25.43 & 27.40
        & 28.85 & 29.82 \\
        \midrule

        \multirow{2}{*}{FedI2Agent}
        & N@10
        & 48.62 & \underline{54.06}
        & \textbf{51.53} & 51.30
        & \underline{51.90} & \underline{49.28} \\
        & MRR
        & 34.52 & \underline{41.44}
        & \textbf{37.22} & 36.55
        & \underline{37.84} & \underline{34.25} \\
        \midrule

        \rowcolor{gray!15}
        & N@10
        & 52.69 & 59.27
        & 50.18 & 54.97
        & \textbf{55.37} & 52.89 \\

        \rowcolor{gray!15}
        \multirow{-2}{*}{\textbf{FedRecAgent$_{D}$}}
        & MRR
        & 38.33 & 46.87
        & 35.42 & 41.09
        & \textbf{41.92} & 38.53 \\
        \midrule

        \rowcolor{gray!15}
        & N@10
        & 51.49 & \textbf{59.73}
        & 49.89 & \textbf{56.58}
        & 52.74 & 54.92 \\

        \rowcolor{gray!15}
        \multirow{-2}{*}{\textbf{FedRecAgent$_{P}$}}
        & MRR
        & 36.91 & \textbf{47.35}
        & 35.20 & \textbf{43.23}
        & 38.59 & 41.07 \\
        \midrule

        \rowcolor{gray!15}
        & N@10
        & \textbf{53.93} & 59.33
        & \underline{50.21} & 55.00
        & 54.61 & \textbf{55.51} \\

        \rowcolor{gray!15}
        \multirow{-2}{*}{\textbf{FedRecAgent$_{P+D}$}}
        & MRR
        & \textbf{39.94} & 46.91
        & \underline{35.53} & 41.04
        & 40.88 & \textbf{41.83} \\

        \bottomrule
    \end{tabular}%
    }
\end{table}

\subsection{Performance Comparison}
Table~\ref{tab:main} reports recommendation performance
on six datasets.
Across all datasets and metrics, at least one
FedRecAgent variant achieves the highest mean.
All three variants outperform the strongest
baselines in N@10 and MRR across all datasets.
The best variants improve by 1.54\%--9.19\% in N@10 and 3.01\%--16.29\% in MRR
over the strongest baselines.
LLM-based baselines generally outperform conventional
federated collaborative filtering and sequential methods,
although the strongest baseline varies across datasets.
We also evaluate FedSASRec-based hard negatives,
ranking each ground-truth item against the nine
highest-scoring non-interacted items 
identified by a trained FedSASRec model.
In Table~\ref{tab:main_hard}, FedRecAgent achieves
the highest N@10 and MRR on five of the six datasets.
The exception is Music-Dense, where FedI2Agent achieves
the best results.
These results support the effectiveness of FedRecAgent
under hard-negative sampling.

\begin{table}[t]
    \centering
    \caption{Ablation Study on FedRecAgent.}
    \label{tab:ablation}
    \footnotesize
    \setlength{\tabcolsep}{2.5pt}
    \renewcommand{\arraystretch}{1.0}
    \setlength{\aboverulesep}{0pt}
    \setlength{\belowrulesep}{0pt}

    \begin{tabular}{@{}c|lcccc@{}}
        \toprule
        \multicolumn{2}{@{}l}{}
        & \multicolumn{2}{c}{CDs-Dense}
        & \multicolumn{2}{c}{Music-Dense} \\
        \cmidrule(lr){3-4}\cmidrule(lr){5-6}
        \multicolumn{2}{@{}l}{Variant}
        & N@10 & MRR & N@10 & MRR \\
        \midrule

        \multicolumn{2}{@{}l}{\textbf{FedRecAgent$_{P+D}$}}
        & \textbf{55.00} & \textbf{41.34}
        & \textbf{53.36} & \textbf{39.39} \\
        \midrule

        \multirow{5}{*}{%
            \rotatebox[origin=c]{90}{%
                \scriptsize\shortstack{Private\\memory}}}
        & w/o $M_u^{\mathrm{sem}}$
        & 50.45 & 35.73 & 48.30 & 33.03 \\
        & w/o $M_u^{\mathrm{dec}}$
        & 52.74 & 38.38 & 52.93 & 38.87 \\
        & w/o $M_u$
        & 46.47 & 30.88 & 47.06 & 31.62 \\
        & w/o evidence extraction
        & 53.52 & 39.55 & 52.17 & 37.90 \\
        & Unified memory
        & 54.07 & 40.19 & 52.42 & 38.09 \\
        \midrule

        \multirow{6}{*}{%
            \rotatebox[origin=c]{90}{%
                \scriptsize\shortstack{Knowledge\\bank}}}
        & w/o shared knowledge
        & 51.45 & 36.85 & 50.08 & 35.22 \\
        & w/o global-guided update
        & 51.76 & 37.39 & 52.58 & 38.34 \\
        & w/o distillation
        & 49.54 & 34.64 & 50.67 & 35.95 \\
        & w/o discrimination
        & 51.48 & 36.78 & 49.13 & 33.97 \\
        & w/o merge
        & 50.53 & 35.80 & 51.42 & 36.84 \\
        & C-only
        & 52.91 & 38.74 & 50.22 & 35.27 \\
        \bottomrule
    \end{tabular}
\end{table}

\subsection{Ablation Study}
\label{sec:ablation}

Table~\ref{tab:ablation} evaluates private memory
and global knowledge construction through component
ablations and alternative designs, with
FedRecAgent$_{P+D}$ as the reference.

\subsubsection{\textbf{Private Memory Construction}}
The private memory group examines the roles of
private memories and episode evidence.
\textbf{w/o $M_u^{\mathrm{sem}}$}: only $M_u^{\mathrm{dec}}$
is used for retrieval, branch selection, and ranking.
\textbf{w/o $M_u^{\mathrm{dec}}$}: only $M_u^{\mathrm{sem}}$
is used for these operations.
\textbf{w/o $M_u$}: the top-$b$ branches with the most assigned knowledge units are shortlisted, and branch selection and ranking proceed without private memories.
\textbf{w/o evidence extraction}: $Z_{u,t}$ is omitted
from reasoning generation and memory updates.
\textbf{w/ Unified memory}: both observations update
one memory, which is used for retrieval,
branch selection, and ranking.
Removing either memory reduces performance, and removing
both causes the largest degradation, supporting their
complementary roles in personalization.
The Unified comparison favors updating
preferences and conditional choice tendencies separately.
Omitting evidence extraction degrades performance,
supporting its role.

\begin{figure*}[t]
    \centering
    \includegraphics[width=1\linewidth]{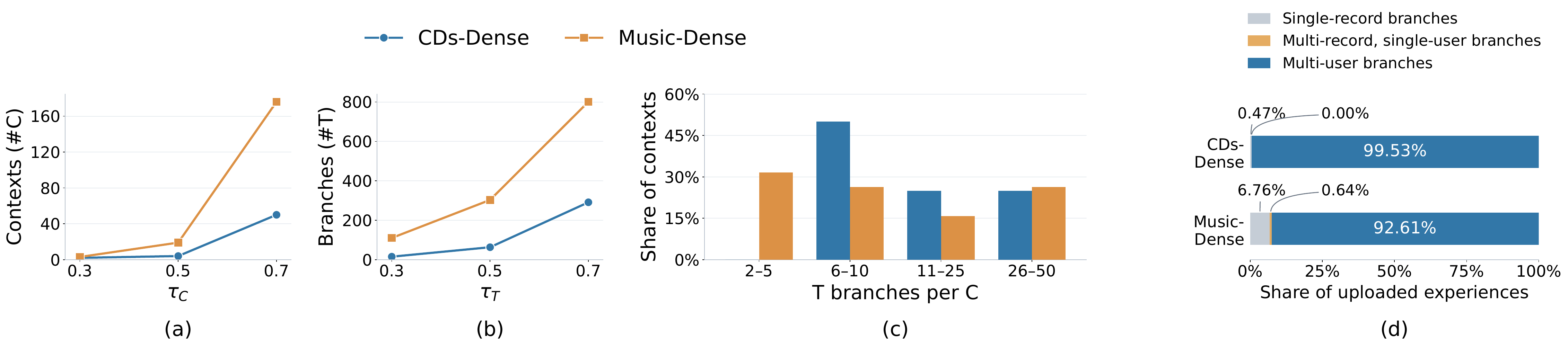}
    \caption{Global knowledge bank structure on CDs-Dense
and Music-Dense.
(a) Condition counts versus $\tau_C$.
(b) Branch counts versus $\tau_T$.
(c) Branches per condition.
(d) Upload shares by branch composition:
single experience, multiple experiences from one user,
or experiences from multiple users.}
    \label{fig:ct_analysis}
\end{figure*}

\subsubsection{\textbf{Global Knowledge Construction}}
This group examines the construction and use of shared
knowledge.
\textbf{w/o shared knowledge}: clients construct private
memories and make recommendations using only their own
interaction histories, without shared knowledge.
\textbf{w/o global-guided update}: clients generate reasoning
without previous shared rules.
\textbf{w/o distillation}: one reasoning record is sampled
per branch instead of consolidating the records.
\textbf{w/o discrimination}: sibling-branch refinement
is omitted.
\textbf{w/o merge}: updated branches use only the current
round's reasoning.
\textbf{w/ C-only}: contributions are grouped by condition
without transition-based partitioning.
All variants reduce performance on both datasets.
Removing shared knowledge highlights the benefit of
cross-user knowledge collaboration.
The remaining comparisons support guiding local reasoning
with shared knowledge, consolidating explanations across
users and rounds, and distinguishing decision transitions.
The C-only comparison further supports condition--transition
alignment for knowledge integration.

\subsection{Study of FedRecAgent}
\label{sec:study}
Additional analyses and qualitative examples are in
Appendix~\ref{appendix:model_analysis}.

\subsubsection{\textbf{Condition--Transition Structure}}
Figure~\ref{fig:ct_analysis} examines the organization
of shared knowledge on CDs-Dense and Music-Dense.
Figures~\ref{fig:ct_analysis}(a) and~\ref{fig:ct_analysis}(b)
vary $\tau_C$ and $\tau_T$, respectively, while fixing
the other threshold at the dataset's selected setting.
Higher thresholds produce finer partitions with more
condition nodes or transition branches.
Figure~\ref{fig:ct_analysis}(c) shows that multiple
transition branches coexist under a single condition,
representing different decision directions under
similar behavioral contexts.
Figure~\ref{fig:ct_analysis}(d) categorizes uploads
according to whether their assigned branch contains
a single experience, multiple experiences from one user,
or experiences from multiple users.
Most uploads belong to multi-user branches,
indicating cross-user knowledge integration.


\begin{table}[t]
    \centering
    \caption{Transition matching accuracy (\%) before and after
    discrimination.
    Scores are averaged within each condition and then across conditions.
    A dash indicates that no eligible branches were available.}
    \label{tab:discrimination_matching}
    \small
    \setlength{\tabcolsep}{4pt}
    \setlength{\aboverulesep}{0pt}
    \setlength{\belowrulesep}{0pt}
    \begin{tabular}{llrrrr}
        \toprule
        Dataset & Stage & R0 & R1 & R2 & R3 \\
        \midrule
        \multirow{2}{*}{CDs-Dense}
        & Before & 45.83 & 55.17 & 79.69 & 73.95 \\
        & After  & \textbf{91.67} & \textbf{79.31}
                 & \textbf{89.06} & \textbf{75.42} \\
        \midrule
        \multirow{2}{*}{Music-Dense}
        & Before & -- & 64.00 & 66.83 & 63.37 \\
        & After  & -- & \textbf{88.00}
                 & \textbf{77.27} & \textbf{75.08} \\
        \bottomrule
    \end{tabular}
\end{table}
\subsubsection{\textbf{Transition Discrimination}}
We assess whether discrimination makes reasoning more
specific to its associated transition.
Given a condition and reasoning statement, Qwen3-8B
selects the corresponding transition from the target
branch and up to four similar sibling branches.
Before--after comparisons use identical candidates
and ordering, with the refinement stage hidden.
Matching accuracy is averaged within each condition
and then across conditions.
Table~\ref{tab:discrimination_matching} shows that
discrimination consistently improves transition matching
across the evaluated dataset--round pairs.
This suggests that refinement makes the applicability
of each branch's reasoning more distinguishable.

\begin{table}[t]
    \centering
    \caption{
    LLM-assessed preservation of preferences and
conditional choice tendencies (\%).
    }
    \label{tab:memory_preservation}
    \small
    \setlength{\tabcolsep}{4pt}
    \setlength{\aboverulesep}{0pt}
    \setlength{\belowrulesep}{0pt}
    \renewcommand{\arraystretch}{1.0}
    \begin{tabular}{@{}lcccc@{}}
        \toprule
        \multirow{2}{*}{Dataset}
        & \multicolumn{2}{c}{Preference}
        & \multicolumn{2}{c}{Tendency} \\
        \cmidrule(lr){2-3}\cmidrule(lr){4-5}
        & Dual & Unified & Dual & Unified \\
        \midrule
        CDs-Dense
        & \textbf{41.59} & 32.27
        & \textbf{26.59} & 16.14 \\
        Music-Dense
        & \textbf{57.47} & 37.22
        & \textbf{43.16} & 26.08 \\
        \bottomrule
    \end{tabular}
\end{table}
\subsubsection{\textbf{Memory Preservation Analysis}}
Table~\ref{tab:memory_preservation} compares separately
maintained semantic and decision memories with unified memory.
Qwen3-8B first extracts preferences and conditional choice
tendencies from training histories without observing
the memories, then evaluates anonymized memories against
these claims.
Claims are classified as preserved, generalized, absent,
or contradicted; preserving a tendency requires retaining
both its context and choice direction.
Preservation rates are averaged across seeds within
each user and then across users.
Dual memory achieves higher preservation rates for both
preferences and choice tendencies on both datasets.
This suggests that separate maintenance helps retain
explicit information about these complementary aspects
of personalization.

\begin{table}[t]
  \centering
  \caption{Recommendation performance (N@10) with different
  client LLM sizes.}
  \label{tab:client_llm}
  \footnotesize
  \setlength{\tabcolsep}{4pt}
  \renewcommand{\arraystretch}{1.05}
  \setlength{\aboverulesep}{0pt}
\setlength{\belowrulesep}{0pt}

  \resizebox{\columnwidth}{!}{%
  \begin{tabular}{@{}lcccccccc@{}}
    \toprule
    \multirow{2}{*}{Method}
    & \multicolumn{4}{c}{CDs-Dense}
    & \multicolumn{4}{c}{Music-Dense} \\
    \cmidrule(lr){2-5}
    \cmidrule(lr){6-9}

    & 0.6B & 1.7B & 4B & 8B
    & 0.6B & 1.7B & 4B & 8B \\
    \midrule

    FedLLMRank
    & 45.82& 44.90& 49.28& 49.91& 47.46& 46.96& \underline{49.14}& 48.67\\

    FedAgentCF
    & 44.75& \underline{46.79}& \underline{50.70}& 50.41& 47.35& \underline{48.65}& 48.78& 47.91\\

    FedAFL
    & \underline{46.11}& 45.57& 49.99& \underline{51.67}& \textbf{47.66}& 47.77& 48.04& \underline{49.31}\\

    FedI2Agent
    & 45.01& 45.16& 46.88& 47.59& 46.79& 45.95& 46.98& 48.75\\

    \midrule

    \cellcolor{gray!15}\textbf{FedRecAgent$_{P+D}$}
    & \cellcolor{gray!15}\textbf{46.25}& \cellcolor{gray!15}\textbf{50.29}& \cellcolor{gray!15}\textbf{55.00}& \cellcolor{gray!15}\textbf{56.81}& 
    \cellcolor{gray!15}\underline{47.64}& \cellcolor{gray!15}\textbf{48.80}& \cellcolor{gray!15}\textbf{53.36}& \cellcolor{gray!15}\textbf{50.43}\\

    \bottomrule
  \end{tabular}%
  }
\end{table}
\subsubsection{\textbf{Client LLM Size}}
Table~\ref{tab:client_llm} compares client LLM sizes.
FedRecAgent consistently achieves the best or second-best
performance across the evaluated configurations.
These results suggest that FedRecAgent remains competitive
across client capacities.


\section{Conclusion}
In this paper, we formulate Federated Recommender Agents, a setting where personalized LLM agents collaborate through natural-language knowledge while keeping private user information local. We propose FedRecAgent, which maintains private dual memories of preferences and choice tendencies, shares C--T--R knowledge units, and hierarchically integrates reasoning by condition and transition. These memories guide personalized retrieval and selection of shared reasoning without updating LLM parameters. Experiments demonstrate improved recommendation performance over baselines, while further analyses support the effectiveness of private memories and structured reasoning integration.

\bibliographystyle{ACM-Reference-Format}
\bibliography{1_ref}
\clearpage
\appendix
\section{Appendix}

\begin{table}[h]
    \centering
    \caption{Dataset statistics.
    Paired values denote Dense/Sparse subsets.
    Density is the interaction count divided by
    the number of possible user--item pairs.}
    \label{tab:dataset_statistics}
    \scriptsize
    \setlength{\tabcolsep}{3pt}
    \renewcommand{\arraystretch}{0.9}
    \begin{tabular}{@{}lrrrr@{}}
        \toprule
        Dataset & Users & Items & Interactions & Density (\%) \\
        \midrule
        CDs \& Vinyl
        & 100 & 1,317/2,681 & 3,117/3,073 & 2.37/1.15 \\
        Musical Instruments
        & 100 & 1,112/2,112 & 2,569/2,867 & 2.31/1.36 \\
        Office Products
        & 100 & 1,092/2,264 & 2,540/2,732 & 2.33/1.21 \\
        \bottomrule
    \end{tabular}
\end{table}

\subsection{Experimental Details}
\label{appendix:exp_details}
\noindent \textbf{Data and Evaluation.}
Following \cite{agentcf,agentcf++,drunkagent}, we construct subsets of CDs \& Vinyl,
Musical Instruments, and Office Products; Table~\ref{tab:dataset_statistics} reports their statistics. Interactions are ordered chronologically within each user, reserving the last two for validation and testing. Training episodes use only the preceding training interactions. For LLM-based methods, items in interaction histories and candidate sets are represented by their titles.

\vspace{+0.1cm}

\noindent \textbf{Sequential Knowledge Updates.}
We process training episodes chronologically to reflect
knowledge accumulation from new experiences.
Unlike repeated local-epoch optimization, our setting
updates natural-language memories and shared knowledge
while keeping LLMs frozen.
Each eligible episode is processed once, with its
information retained through memory updates rather
than repeated replay.
History windows may overlap, but each training interaction
serves as the next-item target only once.
For client coordination, episodes are grouped into
common intervals by target timestamp.
Within each interval, clients process episodes sequentially
using the previous round's shared knowledge, and the
server aggregates their contributions at the end.
Evaluation uses the final memories and knowledge bank.

\vspace{+0.1cm}

\noindent \textbf{Baseline Configurations.}
We compare existing federated recommenders with federated
adaptations of sequential and LLM-based methods, preserving
their core modeling and agent mechanisms within the
client--server setting.
The LLM-based methods are selected for their amenability
to this adaptation.
For conventional and sequential recommenders, we tune
learning rates over $\{10^{-1},10^{-2},10^{-3},10^{-4}\}$
using SGD, with up to 300 communication rounds and ten
local epochs per round.
Sequential models use cross-entropy loss, and other
model-specific settings follow the original implementations.
For conventional and sequential recommenders, we use a batch size of 2048 and full client participation (sampling ratio 1.0). Federated sequential baselines use FedAvg for aggregation.
For LLM-based baselines, FedLLMRank~\cite{llmrank} compares
prompting configurations using interaction history,
recency emphasis, and in-context examples.
FedAgentCF~\cite{agentcf} follows the same chronological
round schedule as FedRecAgent and exchanges only item
memories.
The server LLM merges these memories and broadcasts them
to initialize clients' local item memories for the next round.
Inference combines learned user and item memories with
the ten most recent interactions.
FedAFL~\cite{afl} uses the client LLM for both the recommender
and user simulator, with FedSASRec providing recommendation
priors and reward-model guidance.
Both agents retain the accumulated dialogue history over
up to four feedback iterations, followed by candidate ranking.
FedI2Agent~\cite{i2agent} uses full-history interest memory
and a profile constructed from feedback on a historical
choice, without using held-out test-item feedback.
The client extracts candidate-conditioned interests and
profile information to rerank the initial FedSASRec ranking.

\subsection{Additional Model Analyses}
\label{appendix:model_analysis}

\subsubsection{\textbf{Hyperparameter Study}}
\label{appendix:sensitivity}
\begin{figure}[t]
    \centering
    \includegraphics[width=1.0\linewidth]{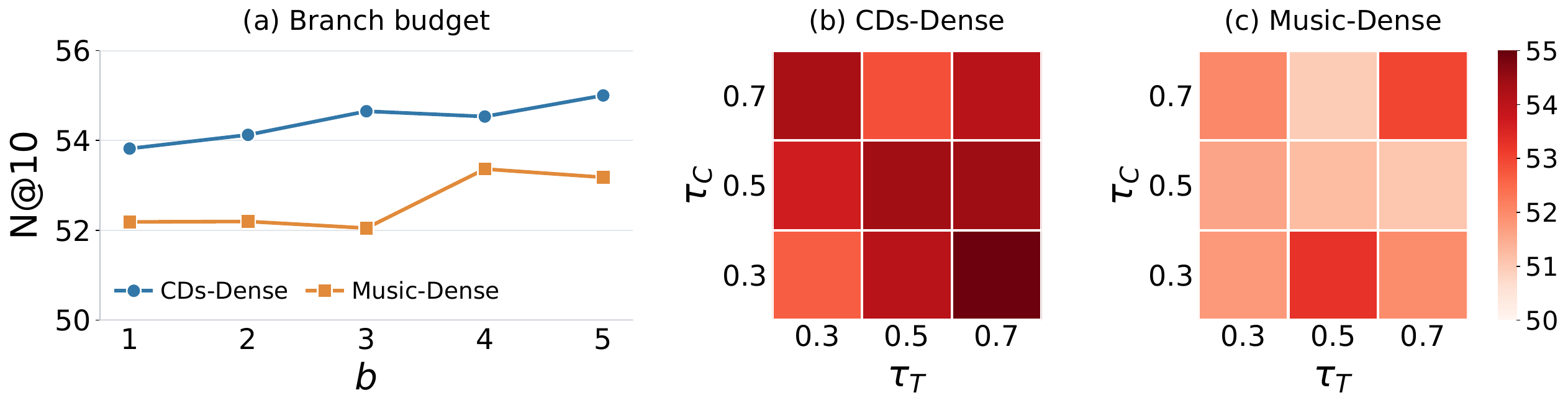}
    \caption{Hyperparameter study on FedRecAgent.}
    \label{fig:sensitivity}
\end{figure}

Figure~\ref{fig:sensitivity} examines the effects of
the branch budget and matching thresholds for
FedRecAgent$_{P+D}$.
Figure~\ref{fig:sensitivity} (a) varies the branch budget $b$, controlling how many retrieved
branches are provided to the selector.
Figure~\ref{fig:sensitivity} (b) and (c) vary the condition and transition
thresholds, $\tau_C$ and $\tau_T$,
which control how knowledge units are grouped in the bank.
Increasing the branch budget does not uniformly improve
performance, and the preferred threshold combinations
differ across datasets.

\begin{table}[t]
    \centering
    \caption{Input-item reconstruction from client uploads,
    measured by Episode Recall@10 (\%).
    Higher values indicate greater reconstruction success.
    Results use 100 users with 10 uploads per user.}
    \label{tab:privacy_reconstruction}
    \scriptsize
    \setlength{\tabcolsep}{10pt}
    \begin{tabular}{lrr}
        \toprule
        Attack input & CDs-Dense & Music-Dense \\
        \midrule
        Random       & 0.71 & 0.88 \\
        Popularity   & 3.85 & 7.87 \\
        \midrule
        $C$          & 0.94 & 1.94 \\
        $T$          & 1.47 & 2.77 \\
        $R$          & 0.80 & 2.68 \\
        $C+T$        & 1.41 & 2.71 \\
        $C+R$        & 0.98 & 2.73 \\
        $T+R$        & 1.20 & 3.15 \\
        \textbf{$C+T+R$} & 1.22 & 3.00 \\
        \bottomrule
    \end{tabular}
\end{table}
\begin{table}[t]
\centering
\caption{
Post-filtered detection of explicit item-title disclosure in
uploaded $(C,T,R)$ records.
Detected entries report the number of uploads and the percentage
of all evaluated uploads.
}
\label{tab:privacy_title_disclosure}
\small
\setlength{\tabcolsep}{7pt}
\begin{tabular}{llrrr}
\toprule
& & & \multicolumn{2}{c}{Detected uploads, $n$ (\%)}\\
\cmidrule(lr){4-5}
Dataset & Uploads/user & Total & 8B & 14B\\
\midrule
\multirow{3}{*}{CDs-Dense}
& 10       & 1,000 & 0 (0.00) & 3 (0.30)\\
& Up to 20 & 1,721 & 0 (0.00) & 5 (0.29)\\
& Up to 30 & 2,117 & 0 (0.00) & 6 (0.28)\\
\midrule
\multirow{3}{*}{Music-Dense}
& 10       & 1,000 & 0 (0.00) & 0 (0.00)\\
& Up to 20 & 1,433 & 0 (0.00) & 0 (0.00)\\
& Up to 30 & 1,569 & 0 (0.00) & 0 (0.00)\\
\bottomrule
\end{tabular}
\end{table}

\subsubsection{\textbf{Input-Item Reconstruction}}
\label{appendix:privacy}
We test whether an attacker can retrieve input items
from a single $(C,T,R)$ upload using the full catalog
of item titles and categories.
Concatenated upload fields are matched to catalog entries
by cosine similarity of \texttt{all-MiniLM-L6-v2} embeddings.
Let $B_e^{10}$ denote the top ten retrieved items and
$S_e$ the unique items in upload $e$'s history and observed
next item.
We measure
$\operatorname{Recall@10}(e)
=|B_e^{10}\cap S_e|/|S_e|$,
averaged over ten uploads per user and then 100 users.
Controls use random retrieval and popularity ranking
by training-item frequency.
Table~\ref{tab:privacy_reconstruction} compares
fields and their combinations.
Complete-record recall exceeds random retrieval, while
all field-based attacks remain below the popularity control.
Uploads thus retain some reconstruction signal.

\subsubsection{\textbf{Explicit Item-Title Detection}}
We inspect nested subsets of 10, up to 20, and up to 30
uploads per user, subject to availability.
Disclosure requires a full input-item title or an identifiable
fragment used as a work name in $(C,T,R)$; generic descriptions,
incidental overlaps, and ambiguous references are excluded.
Independent 8B and 14B LLM judges flag disclosures,
which an AI assistant checks against the uploads and input
titles to remove unsupported matches.
These are post-filtered LLM detections;
each upload is counted once.
Table~\ref{tab:privacy_title_disclosure} reports no confirmed
detections from 8B and rare detections from 14B only on CDs-Dense.
Rates use all evaluated uploads as the denominator.

\begin{figure}[t]
    \centering
    \includegraphics[width=1\linewidth]{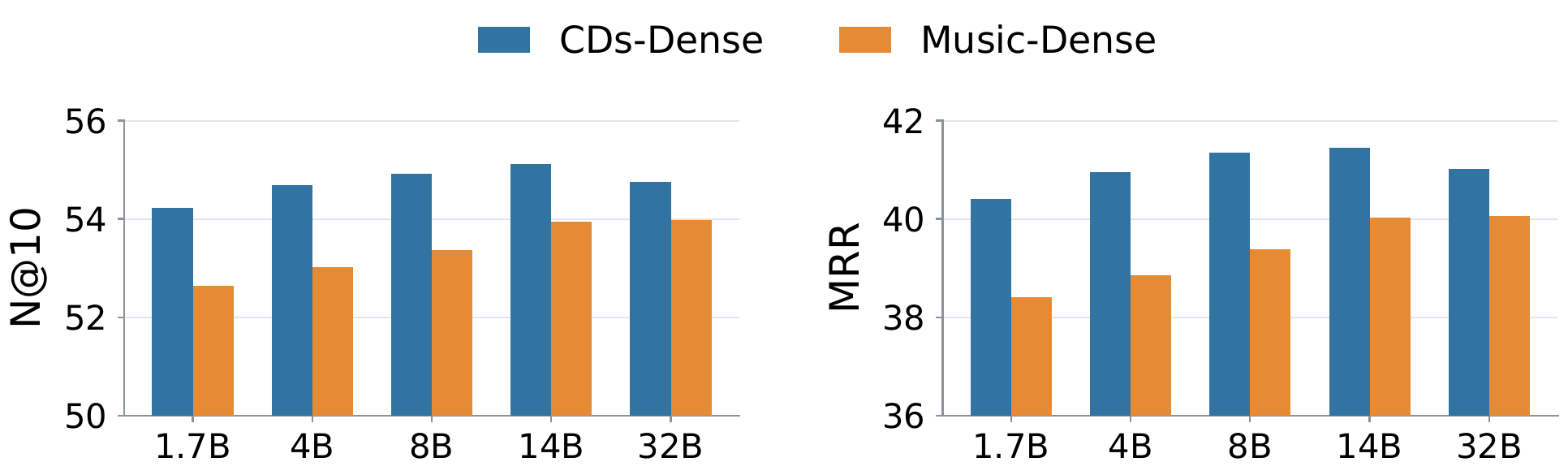}
    \caption{Effect of server LLM size on FedRecAgent's recommendation
performance.}
    \label{fig:server_llm}
\end{figure}
\subsubsection{\textbf{Server LLM Size}}
Figure~\ref{fig:server_llm} compares server LLM sizes
with the client backbone fixed as Qwen3-4B.
Both N@10 and MRR generally improve as server capacity
increases, although the gains are modest and not strictly
monotonic on CDs-Dense.
The default server performs close to the larger configurations,
with only modest additional gains from increased capacity.

\begin{figure}
    \centering
    \includegraphics[width=1\linewidth]{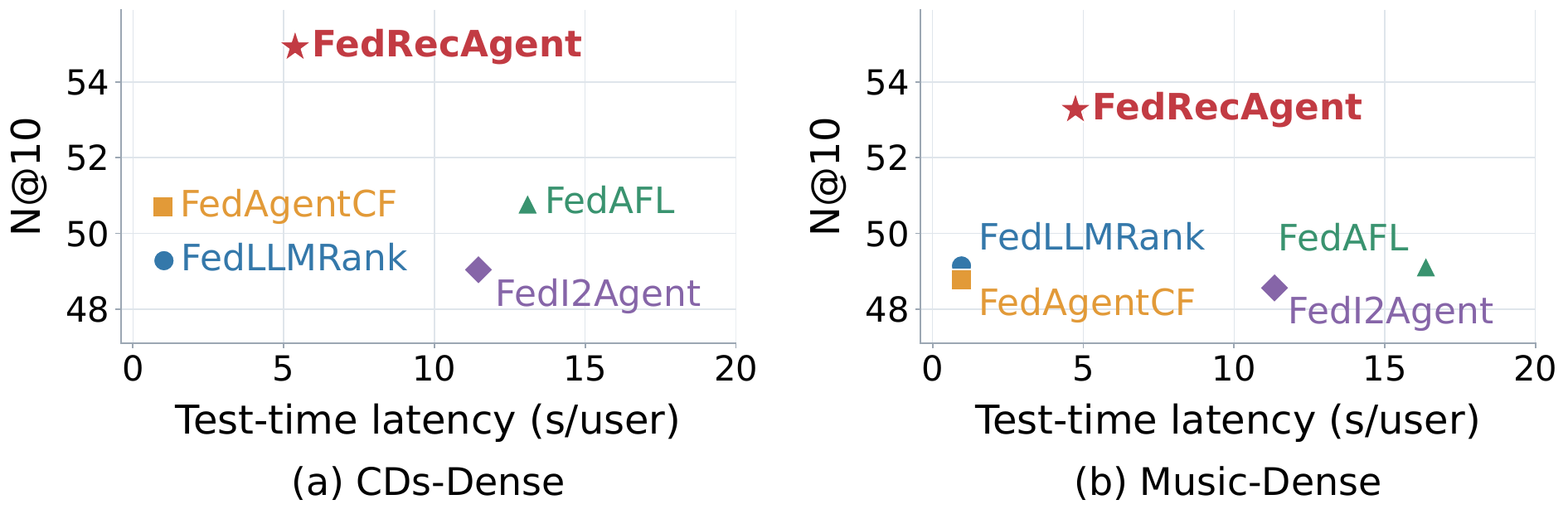}
    \caption{Test-time latency and recommendation performance on CDs-Dense and Music-Dense.}
    \label{fig:time_cost}
\end{figure}
\begin{table}[t]
    \centering
    \caption{
    Client-side storage footprints in decimal units.
Personal state is averaged over users; shared banks
and auxiliary models are counted once per client.
    }
    \label{tab:space_cost}
    \scriptsize
    \setlength{\tabcolsep}{3pt}
    \renewcommand{\arraystretch}{0.9}
    \resizebox{\linewidth}{!}{%
    \begin{tabular}{@{}c|lrrrr@{}}
       \toprule
\multicolumn{1}{@{}c}{}
& Model
& \begin{tabular}[c]{@{}c@{}}Personal state\\(KB/user)\end{tabular}
& \begin{tabular}[c]{@{}c@{}}Shared bank\\(MB)\end{tabular}
& \begin{tabular}[c]{@{}c@{}}Auxiliary model\\(MB)\end{tabular}
& \begin{tabular}[c]{@{}c@{}}Total\\(GB)\end{tabular} \\
\midrule

        \multirow{5}{*}{\rotatebox[origin=c]{90}{\scriptsize CDs-Dense}}
        & FedLLMRank
        & -- & -- & -- & 5.2060 \\
        & FedAgentCF
        & 0.833 & 0.215 & -- & 5.2062 \\
        & FedAFL
        & 1.418 & -- & 0.546 & 5.2065 \\
        & FedI2Agent
        & 2.080 & -- & 0.546 & 5.2065 \\
        & FedRecAgent
        & 1.672 & 0.369 & 91.568 & 5.2979 \\
        \midrule

        \multirow{5}{*}{\rotatebox[origin=c]{90}{\scriptsize Music-Dense}}
        & FedLLMRank
        & -- & -- & -- & 5.2060 \\
        & FedAgentCF
        & 0.856 & 0.218 & -- & 5.2062 \\
        & FedAFL
        & 1.976 & -- & 0.493 & 5.2065 \\
        & FedI2Agent
        & 2.174 & -- & 0.493 & 5.2065 \\
        & FedRecAgent
        & 1.839 & 3.731 & 91.568 & 5.3013 \\
        \bottomrule
    \end{tabular}%
    }
\end{table}
\subsubsection{\textbf{Cost Analysis}}
Figure~\ref{fig:time_cost} compares per-user inference latency
with main-table recommendation performance.
FedRecAgent is slower than FedLLMRank
and FedAgentCF, but faster than FedAFL and FedI2Agent.
It therefore achieves higher recommendation performance
at intermediate inference cost.
Table~\ref{tab:space_cost} compares client-side storage
under a common client LLM backbone.
Storage includes model weights and tokenizer/configuration files. FedAFL and FedI2Agent include serialized request-local state.
FedRecAgent requires additional storage primarily for
the sentence encoder used in local retrieval.
Its personal memories and shared bank occupy relatively
little space compared with the LLM weights.
Although its total footprint is larger than those of
the baselines, model weights remain the dominant
storage cost across all methods.

\begin{figure}[t]
    \centering
    \includegraphics[width=1\linewidth]{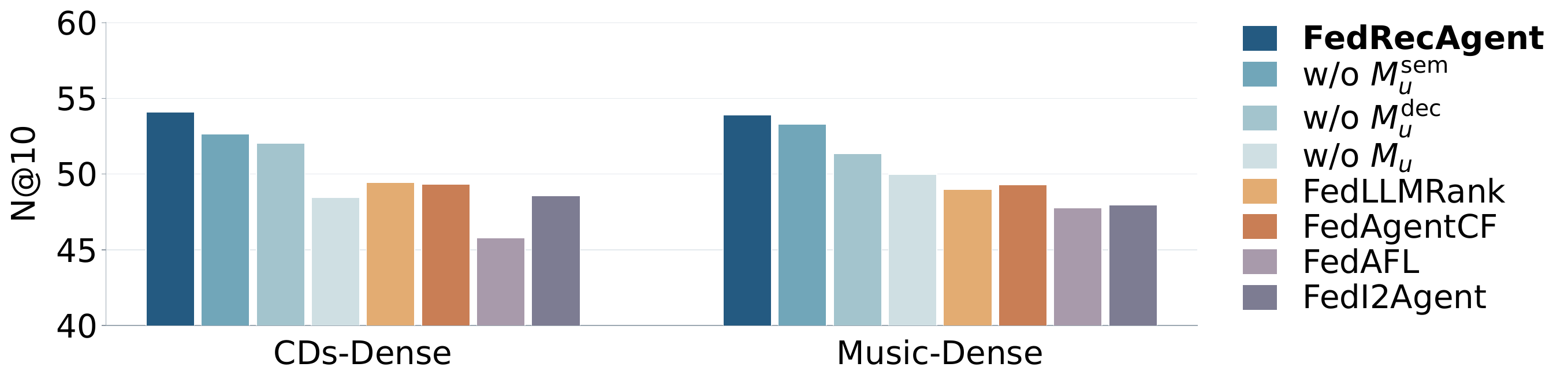}
    \caption{Continual recommendation performance. N@10 is averaged across evaluation blocks
$D_1$--$D_3$.}
    \label{fig:continual}
\end{figure}
\subsubsection{\textbf{Continual Recommendation}}
We process interactions in chronological blocks,
using $D_0$ for initialization and $D_1$--$D_3$ for evaluation.
For each eligible user, the last interaction in each
evaluation block is predicted before observation and
memory updates.
Private memories persist across blocks, while clients
use the previous block's shared bank throughout each block.
Figure~\ref{fig:continual} shows that FedRecAgent achieves
the highest reported N@10 on both datasets.
Removing semantic memory, decision memory, or both
reduces performance, although removing both does not
consistently cause the largest decline.
These results suggest that maintaining personalized
memories supports recommendation as new interactions arrive.

\subsection{\textbf{Qualitative Case Studies}}
\label{appendix:case_studies}
We illustrate personalized memory evolution and two server-side reasoning operations using recorded examples.

\begin{figure}
    \centering
    \includegraphics[width=1\linewidth]{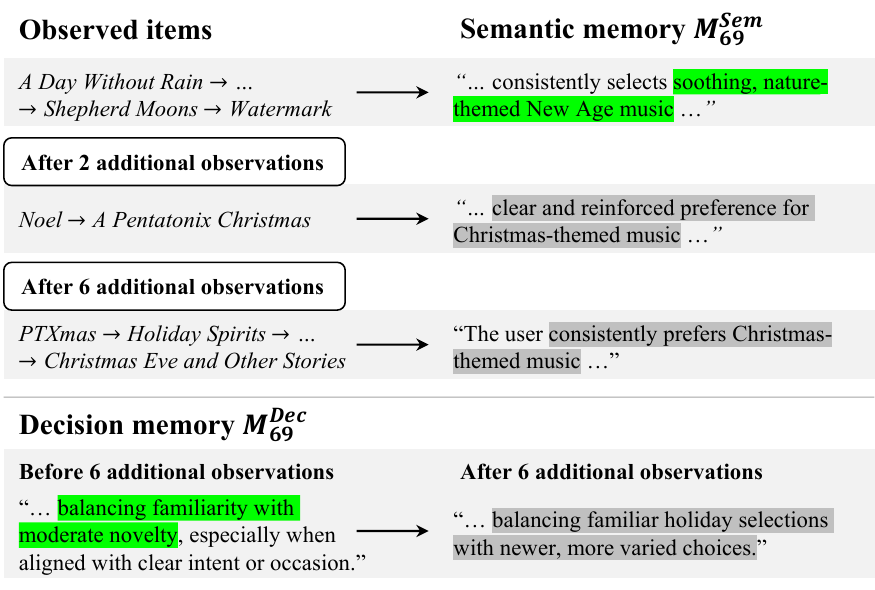}
    \caption{Evolution of personalized memory for User 69 in
CDs-Dense. \colorbox{green}{Green} and
\colorbox{gray!40}{gray} highlight earlier and newly
emphasized content, respectively.}
    \label{fig:case_study_memory}
\end{figure}
\begin{figure}
    \centering
    \includegraphics[width=1\linewidth]{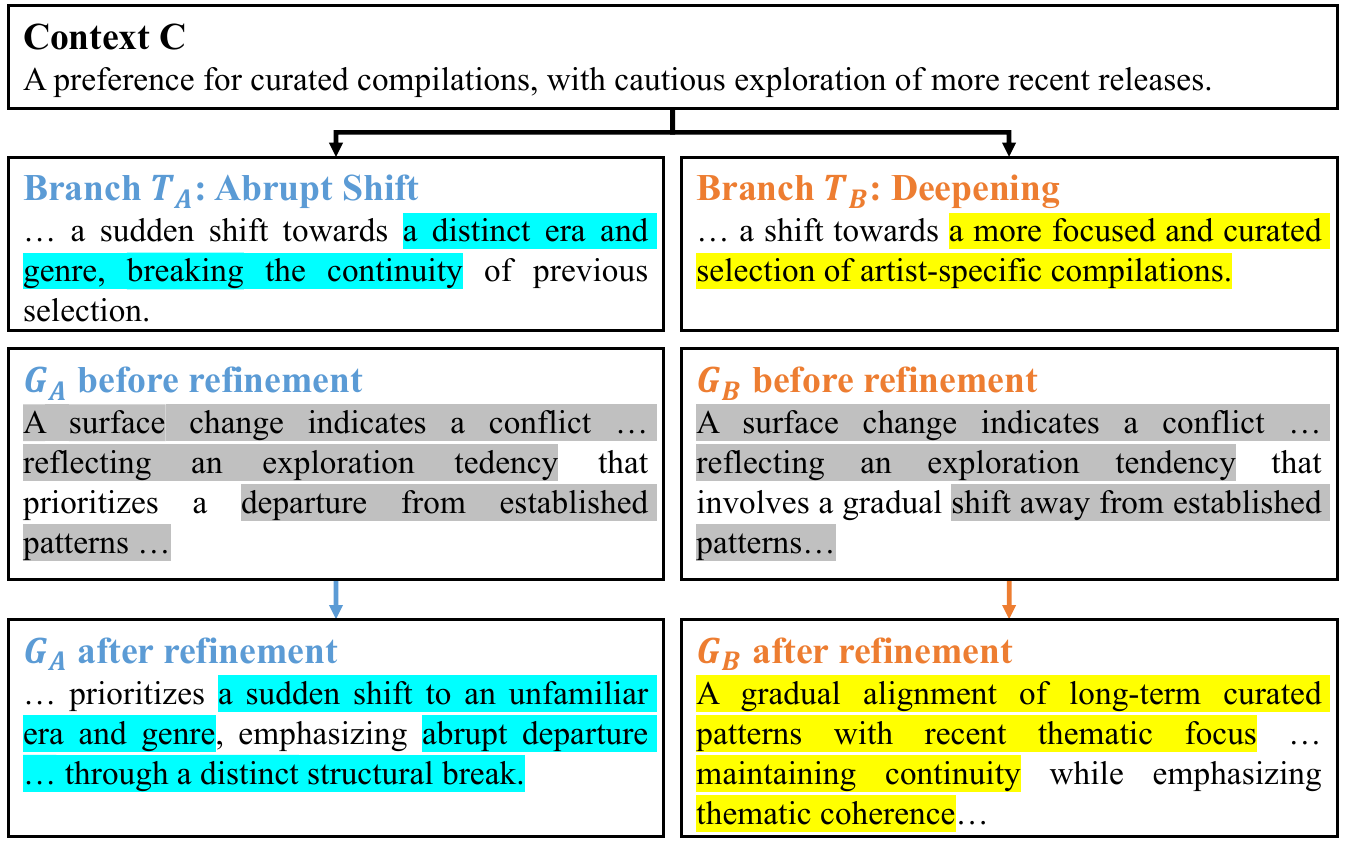}
    \caption{\textbf{Case study of cross-branch discrimination.} 
\colorbox{gray!20}{Gray} highlights overlapping explanations
before refinement; \colorbox{cyan!30}{cyan} and
\colorbox{yellow}{yellow} highlight distinct transition-specific
reasoning afterward.}
\label{fig:case_study}
    \label{fig:case_study_discrimination}
\end{figure}

\subsubsection{\textbf{Evolution of Personalized Memory}}

Figure~\ref{fig:case_study_memory} follows User 69 in CDs-Dense.
After five observations, semantic memory emphasizes soothing,
nature-themed New Age music.
Two subsequent Christmas-themed selections introduce a reinforced
seasonal preference; after six new observations, this interest
becomes the leading emphasis of the memory.
Decision memory also shifts from maintaining calming soundscapes
to balancing familiar holiday selections with newer choices.
This example illustrates shifts in the preferences and
choice tendencies emphasized by the memories.

\subsubsection{\textbf{Cross-Branch Discrimination}}

Figure~\ref{fig:case_study_discrimination} compares two reasoning
statements under a shared condition.
Before discrimination, their descriptions overlap around a
departure from established patterns.
Afterward, one emphasizes an abrupt shift to an unfamiliar era
and genre, while the other emphasizes continuity, thematic
coherence, and artist-specific depth.
The refinement clarifies which transition each explanation
supports without introducing a new observed transition.

\begin{figure}[t]
    \centering
    \includegraphics[width=1\linewidth]{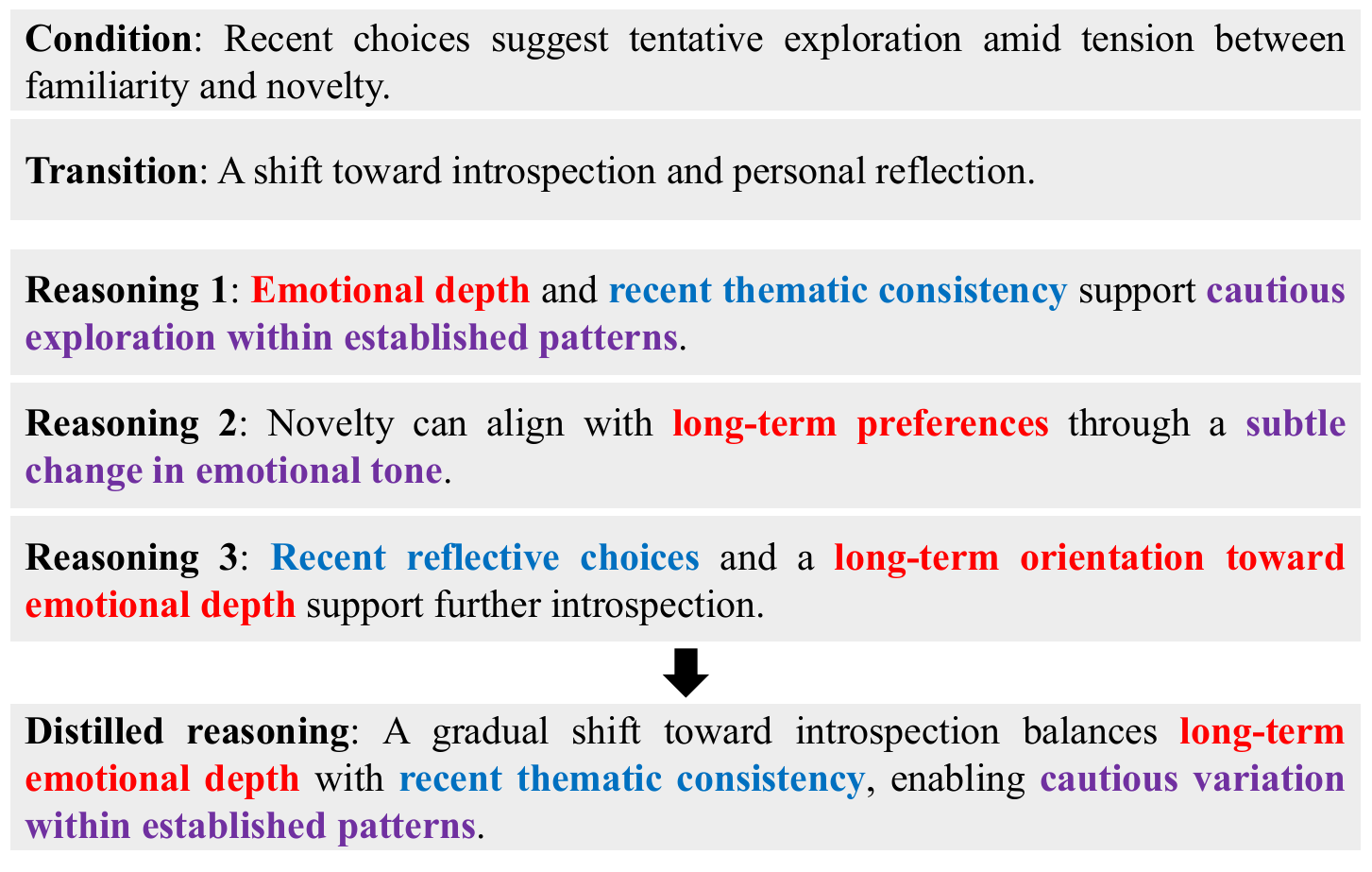}
    \caption{
                A case study of within-branch distillation on CDs-Dense.
                Reasoning from three users assigned to the same
                condition--transition branch is consolidated,
                retaining \redtext{long-term tendencies},
                \bluetext{recent choice signals}, and
                \purpletext{variation within established patterns}.
                Texts are condensed for readability.
                }
    \label{fig:case_study_distillation}
\end{figure}

\subsubsection{\textbf{Reasoning Distillation}}

Figure~\ref{fig:case_study_distillation} illustrates how reasoning
from three users assigned to the same condition--transition
branch is distilled into a shared explanation.
The distilled reasoning connects long-term tendencies
with recent choice signals to explain cautious variation
within established patterns.
This example shows how distillation compresses overlapping
explanations while retaining their shared decision logic.

\end{document}